\documentclass[10pt,letterpaper,twocolumn]{article} 

\usepackage[top=0.75in,bottom=0.75in,left=0.75in,right=0.75in]{geometry}

\usepackage{setspace}
\usepackage[utf8]{inputenc}
\usepackage[utf8]{inputenc}
\usepackage{newunicodechar}
\newunicodechar{́}{\'}
\usepackage[backend=bibtex, style=numeric-comp, sorting=none]{biblatex}
\usepackage{nameref,hyperref}

\usepackage{microtype}
\DisableLigatures[f]{encoding = *, family = * }

\usepackage{indentfirst} 
\usepackage{titlesec}
\usepackage{amssymb}
\titleformat{\section}{\normalsize\bfseries}{\thesection}{1em}{}
\titleformat{\subsection}[runin]{\small\bfseries}{\thesubsection}{1em}{}[.]

\titlespacing*{\section}{0pt}{\baselineskip}{1em}
\titlespacing*{\subsection}{0pt}{0.5\baselineskip}{1em}

\usepackage{changepage}

\usepackage[font=small,aboveskip=1pt,labelfont=bf,labelsep=period,singlelinecheck=off]{caption}

\usepackage{lastpage,fancyhdr,graphicx}
\usepackage{epstopdf}
\fancyheadoffset[L]{0in}
\fancyfootoffset[L]{0in}

\usepackage{color}

\definecolor{Gray}{gray}{.25}

\usepackage{graphicx}

\usepackage{sidecap}

\usepackage{wrapfig}
\usepackage[pscoord]{eso-pic}
\usepackage[fulladjust]{marginnote}
\reversemarginpar

\usepackage{amsmath}
\usepackage{booktabs}
\usepackage{placeins}

\newcounter{movie}
\def\movie#1{%
  \stepcounter{movie}%
  \noindent\textbf{Movie S\arabic{movie}.} #1\par\vspace{0.5em}%
}

\newcounter{dataset}
\def\dataset#1#2{%
  \stepcounter{dataset}%
  \noindent\textbf{Dataset S\arabic{dataset}.} (#1) #2\par\vspace{0.5em}%
}

\begin{document}

\begin{refsection}

\twocolumn[
  \begin{center}
    {\Large\textbf{A developmental switch from capillary rectification to elastic catapult enables honeydew ejection in the spotted lanternfly}}
    \vspace{0.5cm}
    
    \begin{small}
    Nami Ha$^{1,2}$, Elio J. Challita$^{1,2}$, Jacob S. Harrison$^{1,3}$, Elizabeth G. Clark$^{4,5}$, Kendall E. Larson$^{1}$, Miriam F. Cooperband$^{6}$, \\ Saad Bhamla$^{1,7,*}$
    \end{small}
    \vspace{0.3cm}
    
    \begin{footnotesize}
    $^1$ School of Chemical and Biomolecular Engineering, Georgia Institute of Technology, $^2$ George W. Woodruff School of Mechanical Engineering, Georgia Institute of Technology, $^3$ Department of Biology, Morehouse College, $^4$ Lawrence Berkeley National Laboratory, \\
    $^5$ Department of Environmental Science, Policy and Management, University of California Berkeley, $^6$ Forest Pest Methods Laboratory, S\&T-PPQ-APHIS, United States Department of Agriculture, $^7$ University of Colorado Boulder, BioFrontiers Institute and Department of Chemical and Biological Engineering, Colorado, USA \\
    * saad.bhamla@colorado.edu
    \end{footnotesize}
    \vspace{0.5cm}
  \end{center}
    
  \noindent \textbf{Abstract} 
  
  \noindent Plant sap-feeding insects must dispose of excess fluid, yet at millimeter scales droplet release is constrained by capillary adhesion and contact-line pinning. How phloem-feeding insects solve this puzzle, particularly as the excretory apparatus changes in size and form from nymph to adult, has remained unclear. Combining micro-CT, high-speed imaging, measurements of honeydew properties, and reduced-order modeling, we show that the spotted lanternfly (\textit{Lycorma delicatula}) uses distinct release mechanics across ontogeny. Nymphs release honeydew with an anal stylus that acts as a capillary rectifier, imposing a curvature asymmetry that biases the attached droplet toward detachment through a Laplace-pressure difference. Adults use a longer stylus associated with an elastic basal region, maintain stylus--droplet contact through a finite compression phase, and release droplets with greater translational and rotational momentum. In both stages, stylus rotation is ultrafast, with peak angular accelerations of order $10^7$~rad\,s$^{-2}$ and release unfolding on millisecond timescales, yet droplet ejection speed remains below stylus tip speed. Weber--Bond scaling based on measured honeydew properties places both stages at $We_d<1$ and $Bo_d<1$ at the outlet, but distinguishes their post-release states: nymphal droplets remain surface-tension dominated, whereas adult droplets enter deformation- and spin-influenced regimes. Development therefore maintains waste clearance across ontogeny under the same outlet-scale capillary constraint by changing how stylus motion is coupled to the droplet at release, linking life-stage biomechanics to honeydew placement in this invasive phloem feeder and suggesting bioinspired strategies for droplet ejection, antifouling, and self-cleaning surfaces.
  \par
  \vspace{1cm}
]

\section*{Significance}
\vspace{-0.5\baselineskip}
Small sap-feeding insects must remove waste droplets under strong capillary adhesion, yet how this constraint is met over development has remained unclear. We show that the spotted lanternfly solves this capillary constraint with two distinct release strategies: capillary rectification in nymphs and a subpropulsive, droplet-launcher in adults. This developmental shift connects droplet-ejection mechanics to the biology of an invasive phloem feeder whose honeydew fouls plants and surfaces and contributes to aggregation-related chemical cues. By linking fluid mechanics, life-stage biomechanics, and honeydew properties, this work could guide multi-modal lure design and infestation management, while also identifying bioinspired principles for small-scale droplet ejection and self-cleaning surfaces.

\begin{figure*}[t!]
\centering
\includegraphics[scale=0.5]{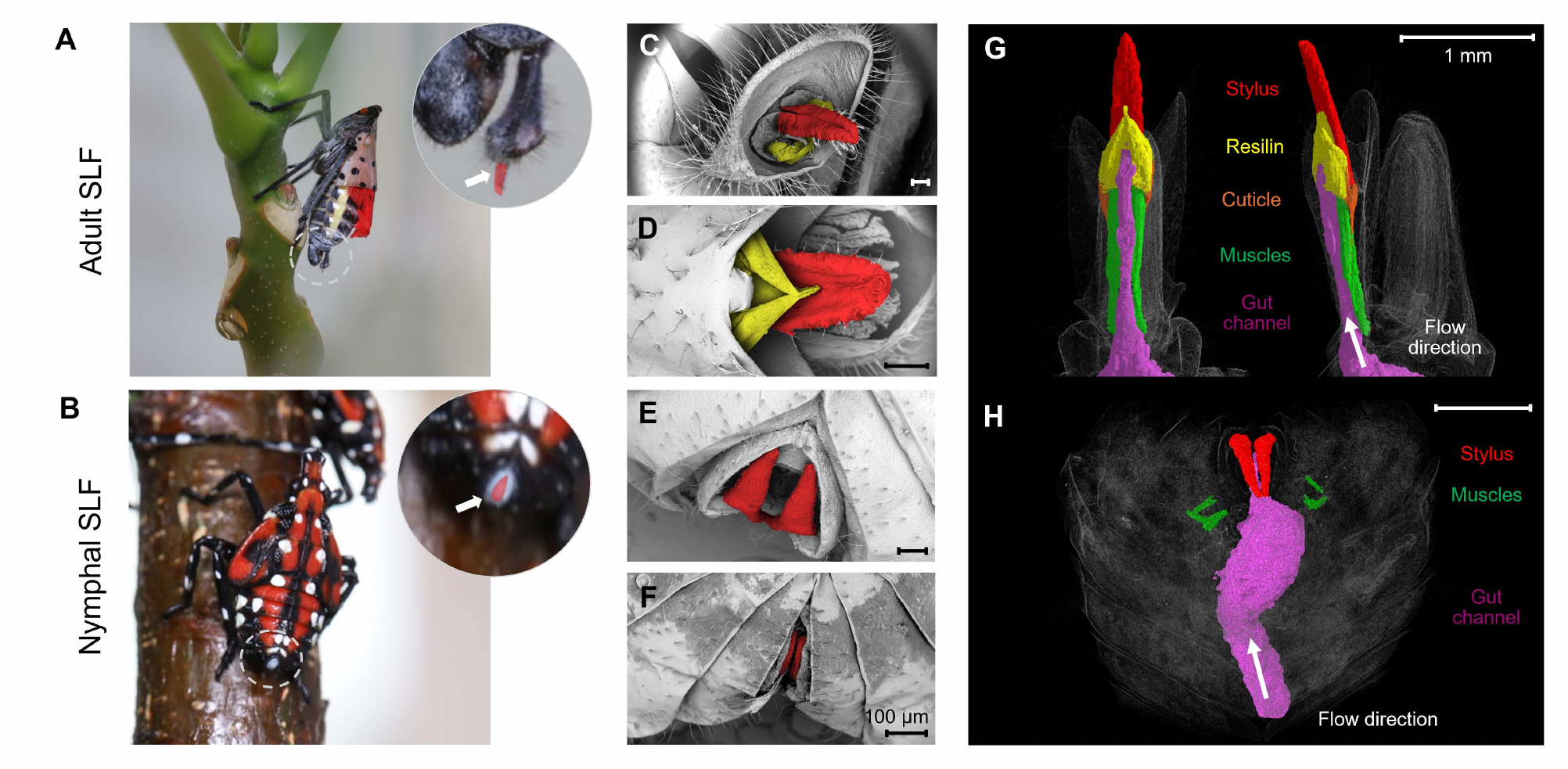}
\caption{\textbf{Developmental remodeling of the anal stylus from nymph to adult spotted lanternfly (SLF, \textit{Lycorma delicatula}).} (A,B) Adult male and 4th instar nymph feeding on phloem sap; arrows mark the anal stylus. (C-F) SEM images of adult (C-D) and nymphal (E-F) styli. Both (C) female and (D) male adults show a distinct basal joint region consistent with a resilin-rich elastic element, whereas nymphs (E:4th instar and F:1st instar) lack an obvious basal elastic region. (G,H) Micro-CT reconstructions of the terminal abdomen of a male adult (G) and 4th instar nymph (H); the stylus and resilin are highlighted (red and yellow), respectively.}
\label{figure_1}
\end{figure*} 

\section*{Introduction}
\vspace{-0.5\baselineskip}
Sap-feeding insects ingest large volumes of plant sap and must expel the excess continuously to maintain osmotic balance and prevent self-fouling~\cite{problemsofphloemfeeding2006angela, osmoticpotential2011julien}. For these insects, excreted droplets are typically in the sub-millimeter to millimeter range, where surface tension and contact-line pinning can exceed gravity and strongly resist detachment, so a droplet can remain attached even when it is comparable in size to the insect. Waste clearance, therefore, becomes a coupled fluid--structure interaction problem in which appendage geometry, actuation, and fluid properties set the force balance between capillarity, inertia, and viscous dissipation.

Different sap feeders meet this constraint with distinct mechanical and materials strategies. Xylem sap-feeding sharpshooters (\textit{Cicadellidae}) form discrete water-like droplets on an anal stylus and then catapult them, leveraging droplet elasticity and tuning stylus motion to the droplet’s inertio-capillary response (``superpropulsion'')~\cite{sharpshooter2023elio,Fluidejection2024elio}. Cicadas (\textit{Cicadidae}) process large volumes by expelling jets~\cite{cicada2024elio}, whereas spittlebug nymphs (\textit{Cercopoidea}) stabilize excreta as foam~\cite{spittlebug2024hannelore,Fluidejection2024elio}. Phloem feeders add further layers of control: aphids can kick honeydew away or coat droplets with hydrophobic wax to form rolling ``liquid marbles,'' reducing adhesion and coalescence~\cite{aphid2002nathan, liquidmarble2018pierre}. Together, these cases show that appendage geometry, actuation, and wetting properties shift the balance among capillarity, inertia, viscosity, and gravity to enable droplet detachment and transport. Less clear is how these detachment and launch reorganize across development, as the excretory apparatus changes size and geometry while capillary constraints persist.

Phloem sap is enriched in sugars and other solutes~\cite{phloemsapchemcomp1972Shelagh, phloemsugarconcent2013kaare, phloemxylemcompare1983Elisabeth}, and excess carbohydrates are excreted as honeydew~\cite{aphid2002nathan, SLF2015surendra, SLF2022hajar}. Honeydew can wet and contaminate the body and surrounding plant surfaces, promoting microbial and fungal growth and altering ecological interactions~\cite{aphid2002nathan, SLF2015surendra, SLF2022hajar, lanternbug2007piotr}. Ballistic honeydew ejection has been reported in fulgorid lanternbugs (e.g., \textit{Enchophora} and \textit{Phrictus}) at speeds approaching order 1 m/s, but the physical mechanism remains unresolved~\cite{lanternbug2007piotr}. This raises three questions: how are droplets detached despite capillary adhesion and pinning, how is impulse transferred into the droplet at release, and how these strategies change as the excretion apparatus grows across development?

Here, we investigate the honeydew ejection mechanism in the spotted lanternfly (SLF, $\textit{Lycorma delicatula}$), a severely damaging invasive phloem feeder that produces copious honeydew~\cite{SLF2015surendra, SLF2022hajar,SLFlure2024miriam,SLFvibroacoustic2022barukh}. We combine micro-CT and microscopy to quantify the anal stylus and surrounding structures, high-speed imaging to resolve droplet growth and release, and mathematical modeling to elucidate the underlying force balances. We show that SLF reorganizes the coupling between stylus motion and droplet detachment across development: nymphs use a hinge-like stylus that rectifies capillary forces to shed droplets with minimal droplet--stylus contact time, whereas adults develop an elastic element that loads during droplet growth and releases rapidly to catapult droplets into an inertia-influenced deformation regime. These results provide a mechanistic explanation for how morphological growth can drive a stage-specific switch between capillary and elastic strategies while operating under invariant surface-tension constraints.

\begin{figure*}[t!]
\centering
\includegraphics[scale=0.5]{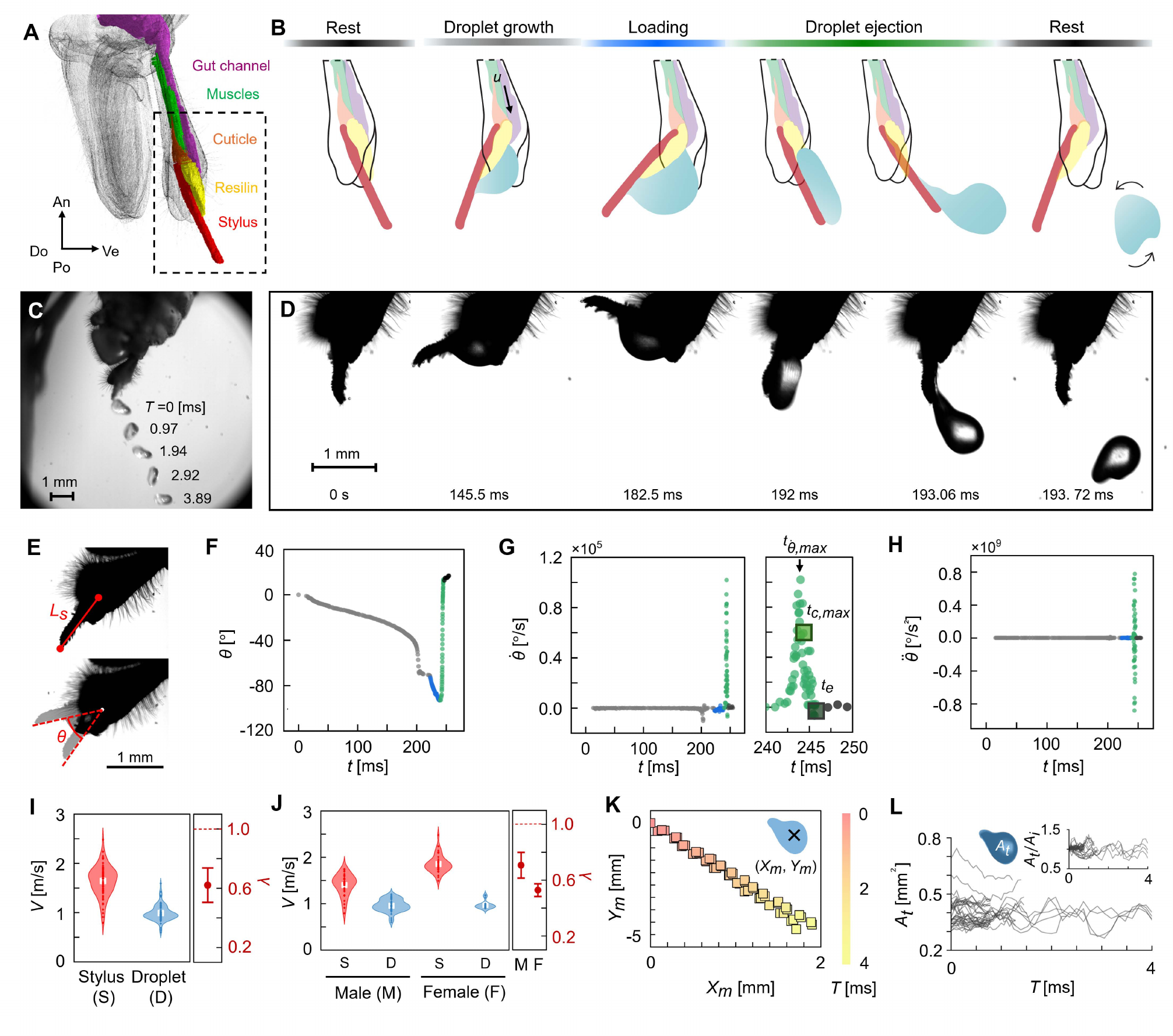}
\caption{\textbf{Subpropulsive honeydew ejection in adult spotted lanternflies (SLF) is mediated by an elastic stylus catapult.} (A) Micro-CT reconstruction of the terminal abdomen of an adult SLF showing the anal stylus and its orientation relative to the body axes. (B) Schematic of the ejection sequence: droplet growth, loading of an elastic element as the stylus deflects, and rapid release that launches the droplet. (C) Overlay of post-launch droplet shapes over time illustrating inertia-influenced deformation. (D) Representative high-speed image sequence of an adult ejection event. (E) Definition of stylus length $L_s$ and stylus angle $\theta$ measured about the pivot point. (F-H) Stylus kinematics: (F) stylus angle $\theta(t)$, (G) angular velocity of the stylus $\dot{\theta}(t)$, and (H) angular acceleration of the stylus $\ddot{\theta}(t)$. Maximum droplet compression ($t_\textit{c, max}$), peak angular speed ($t_{\dot{\theta},  \textit{max}}$), and droplet ejection ($t_e$) occur in rapid succession (within $\sim$1~ms). (I) Droplet launch speed $V_d$ is lower than stylus tip speed $V_s$, giving a speed ratio $\lambda=V_d/V_s<1$ (subpropulsion~\cite{sharpshooter2023elio}) ($V_d$: $n=10$ individuals, $N=84$ ejections; $V_s$: $n=6$, $N=70$). (J) Subpropulsion ($\lambda<1$) occurs for both male and female adults. Error bars show mean $\pm$ one standard deviation; and white markers denote means. (K) Early-time center-of-mass trajectories of ejected honeydew droplets in the $X-Y$ plane; droplet position is $(X_m,Y_m)$. (L) Post-launch deformation quantified by droplet surface area $A_t$ and normalized surface area $A_t/A_i$ (inset), where $A_i$ is the surface area at ejection ($T=0\equiv t_e$) ($n=6$, $N=34$). }
\label{figure_2}
\end{figure*}

\section*{Results}
\vspace{-0.5\baselineskip}
\subsection*{Developmental transition of stylus and surrounding appendages from nymph to adult SLF}
Micro-CT and SEM imaging reveal a marked reorganization of the anal stylus and distal gut outlet between nymphal and adult stages of SLF (Figure~\ref{figure_1}, Figure~S1, Supplementary Video~1). In adults, the stylus forms a slender, elongated appendage with length of $L_\textit{adult}=$1.2$\pm$0.3~mm (2 male and 2 female CT scans) and is anchored at the terminal abdomen by a compliant resilin basal region consistent with an elastic spring (Figure~\ref{figure_1}C--D,G) \cite{sharpshooter2023elio, resilin_2008_malcolm, SLFresilin_2025_Xu}. Across four micro-CT reconstructions (2 male, 2 female), the distal gut channel narrows to an effective diameter $d$= 0.052$\pm$0.007~mm (Figure \ref{figure_1}G, Figure S1). Longitudinal muscle bundles align with and surround the proximal gut channel near the stylus base (Figure \ref{figure_1}G, Supplementary Video~1). 

In contrast, nymphs possess a shorter stylus composed of two closely apposed lobes that form a hinge-like geometry (Figure~\ref{figure_1}E-F,H). For fourth instars, the stylus length is $L_\textit{nymph}=$0.7~mm (2 CT scans, 0.6 mm and 0.8 mm), and the stylus morphology is conserved from the first to fourth instar (Figure~\ref{figure_1}E-F, Figure~S1). We did not observe an obvious basal compliant resilin element analogous to the adult structure (Figure~\ref{figure_1}E,H). The effective distal gut channel diameter is substantially wider $d$=0.336$\pm$0.047~mm (2 micro-CT scans and one SEM measurement). Muscle bundles are present near the terminal abdomen, but adult-like longitudinal bundles wrapping the proximal gut channel are not evident at the resolution of our nymph scans. 

Together, the modest increase in stylus length, the $\sim$6-fold narrowing of the outlet, and the emergence of a compliant basal region and distinct musculature in adults indicate a developmental reorganization of the ejection apparatus that should alter how stylus motion couples to droplet detachment and launch.

\begin{figure*}[t!]
\centering
\includegraphics[scale=0.5]{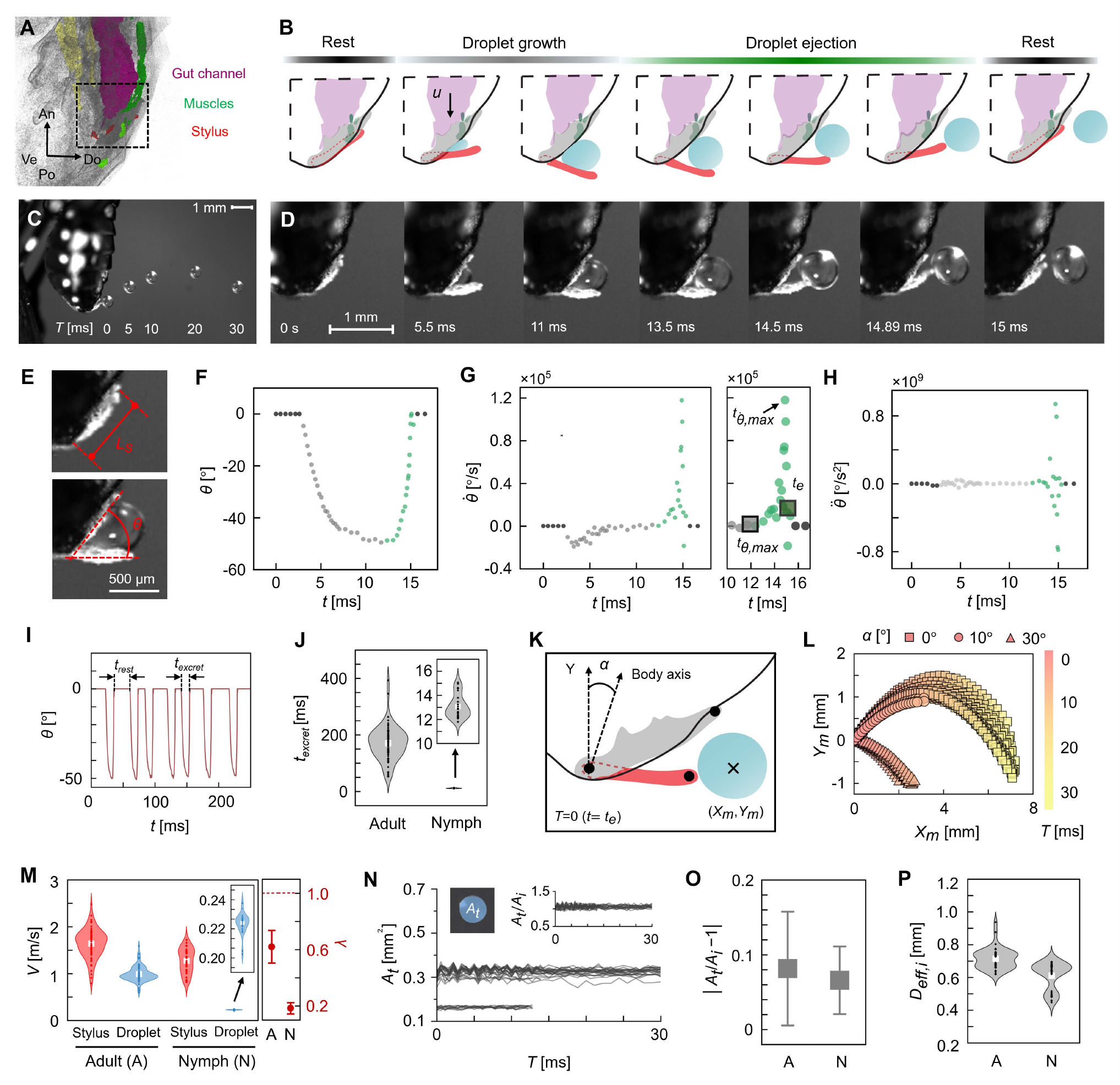}
\caption{\textbf{Honeydew ejection in nymphal SLF is mediated by a hinge-like stylus.} (A) Micro-CT reconstruction of the terminal abdomen showing the anal stylus and its orientation relative to the body axes. (B) Schematic of the nymphal ejection sequence, consisting of droplet growth followed by stylus closure and droplet ejection. (C) Representative post-launch droplets images over time. (D) High-speed image sequence of a nymphal ejection event. (E) Definition of stylus length $L_s$ and stylus angle $\theta$ measured about the pivot point. (F-H) Stylus kinematics: (F) $\theta(t)$, (G) $\dot{\theta}(t)$, and (H) $\ddot{\theta}(t)$, showing that timing of maximum angular displacement ($t_{\theta,\textit{max}}$), maximum angular speed ($t_{\dot{\theta}, \textit{max}}$), and droplet ejection ($t_e$) occur sequentially within $\sim$3~ms. (I) Sequential ejection timing: rest time $t_\text{rest}$ (stylus closed; $\theta=0^\circ$) and excretion time $t_\text{excret}$ (stylus open; $\theta<0^\circ$). (J) Excretion time $t_{\mathrm{excret}}$ of nymphal SLF is much shorter than adult SLF. (K) Definition of body orientation $\alpha$ relative to the longitudinal body axis. (L) Early-time center-of-mass trajectories of droplets in the $X$--$Y$ plane shown as a function of $\alpha$. (M) Droplet speed is lower than the stylus tip speed, as quantified by $\lambda=V_d/V_s<1$ for nymphs and adults. (N) Temporal variation of droplet projected area $A_t$ for nymphal SLF ($n=3$, $N=36$); inset shows normalized area $A_t/A_i$ with $T=0\equiv t_e$. (O) Mean deviation $|A_t/A_i-1|$ comparing adult droplets (A, Figure.~2L) and nymph droplets (N). (P) Effective diameter at ejection $D_{\mathrm{eff,i}}$ compared across adults and nymphs. Error bars show mean $\pm$ one standard deviation; white markers denote means.}
\label{figure_3}
\end{figure*}

\begin{table*}[t!]
\centering
\footnotesize 
\caption{Summary of droplet and stylus kinematics for nymphal and adult SLF. Values are mean $\pm$ SD where measured.}
\begin{tabular*}{\textwidth}{@{\extracolsep{\fill}}lllccc@{}}
\toprule
System & Parameter & Description & Unit & Nymph & Adult  \\
\midrule
Droplet 
& $m_d$ 
& Mass 
& kg
& $1.2 \times10^{-7}$ 
& $1.8 \times 10^{-7}$ \\

& $A_t$ 
& Projected area 
& mm$^2$
& $0.30 \pm 0.06$ 
& $0.40 \pm 0.07$ \\

& $|A_t/A_i - 1|$ 
& Projected-area fluctuation 
& ---
& $0.06 \pm 0.05$ 
& $0.09 \pm 0.08$ \\

& $D_{\mathrm{eff,i}}$ 
& Initial effective diameter 
& mm
& $0.61 \pm 0.07$ 
& $0.71 \pm 0.06$ \\

& $|D_{\mathrm{eff},t}/D_{\mathrm{eff},i}-1|$ 
& Diameter fluctuation 
& ---
& $0.03 \pm 0.02$ 
& $0.04 \pm 0.04$ \\

& $V_{d}$ 
& Ejection velocity 
& m/s
& $0.22 \pm 0.01$
& $0.96 \pm 0.12$ \\

& $\Omega$ 
& Post-launch Angular velocity 
& rad/s
& $\sim 8.06 \times 10^{2}$ 
& $\sim 2.91 \times 10^{3}$ \\

& $E_K$ 
& Translational kinetic energy 
& J
& $\sim 2.90 \times 10^{-9}$ 
& $\sim 8.64 \times 10^{-8}$ \\

& $E_R$ 
& Rotational kinetic energy 
& J
& $\sim 1.45 \times 10^{-9}$ 
& $\sim 3.84 \times 10^{-8}$ \\

\midrule
Stylus 
& $L_s$ 
& Average stylus length 
& mm
& 0.7 (micro-CT data) 
& 1.1 (micro-CT data) \\

& $V_{s,\max}$ 
& Maximum velocity 
& m/s
& $1.27 \pm 0.26$ 
& $1.64 \pm 0.31$ \\

& $a_{s,\max}$ 
& Maximum acceleration 
& m/s$^2$
& $11{,}957 \pm 3{,}861$ ($\sim 1220\,g$) 
& $9{,}214 \pm 3{,}974$ ($\sim 940\,g$) \\
\bottomrule 
\end{tabular*}
\label{table3}
\end{table*}

\subsection*{Droplet ejection by adult SLF using a catapult-like stylus}
The honeydew ejection process of adult SLF consists of three phases: droplet growth, spring loading, and droplet ejection (Figure~\ref{figure_2}A-D, Supplementary Video~2). During droplet growth, honeydew accumulates at the distal end of the stylus (Figure~\ref{figure_2}B,D). Near the end of the growth, we observe a metastable interval during which the stylus angle $\theta$ remains nearly constant while the droplet continues to grow (Figure~S2). This interval lasts $t_m=$17~ms on average, with similar durations in male and female SLF (19~ms and 16~ms on average, respectively, Figure~S2). The event then enters a loading phase in which the stylus rotates further, consistent with progressive deformation (energy loading) of the basal elastic element (Figure~\ref{figure_2}B,D). Finally, rapid release compresses the droplet and launches it from the stylus tip. 

Using stylus length and angular position (Figure~\ref{figure_2}E), we quantify ejection kinematics. A full ejection event lasts $\sim$200--250~ms and is dominated by droplet growth (Figure \ref{figure_2}D,F, Supplementary Video 2). During the brief ejection phase, the stylus undergoes an ultrafast rotation with peak angular speed $\dot\theta_\textit{max}\sim10^5$~$^\circ$/s (Figure~\ref{figure_2}G). Peak angular speed is followed by maximum droplet compression ($t_{c,\max}$) and droplet separation ($t_e$) within $\sim$1--2~ms. Peak angular acceleration reaches $\ddot{\theta}_{\max}\sim 10^{9}$~$^\circ/$s$^{2}$ (Figure.~\ref{figure_2}H), corresponding to maximum tip accelerations on the order of $L_\mathrm{adult}\ddot{\theta}_{\max}\sim 10^{4}$~m/s$^{2}$ ($\sim1000\space g$). 

Despite the ultrafast stylus motion, droplet launch remains subpropulsive: droplet speed ($V_d$) is lower than the maximum stylus tip speed ($V_s$) (Figure~\ref{figure_2}I)~\cite{sharpshooter2023elio}. We define subpropulsive ejection as $V_d<V_s$, or equivalently $\lambda=V_d/V_s<1$. Across ejection events, $V_d=0.96\pm0.12$~m/s and $V_s=1.64\pm0.31$~m/s, yielding $\lambda=0.62\pm0.11$. Subpropulsion is observed for both sexes, with $\lambda=0.71\pm0.09$ (male; n=3, N=30) and $\lambda=0.53\pm0.04$ (female; n=3, N=28) (Figure~\ref{figure_2}J).

After launch, adult honeydew droplets deform and rotate in flight (Figure~\ref{figure_2}C,L; Supplementary Video~3). Over the first 3--4~ms, gravity produces little deflection on the scale of the observed motion, and trajectories appear approximately linear within the imaging window (Figure~\ref{figure_2}C,K). Droplets translate on the order of a millimeter within $\sim$1.5~ms (n=6, N=33; Figure~\ref{figure_2}K). The deformation is substantial as the droplet projected area fluctuates by $\sim$10\% relative to its value at ejection (mean $|A_t/A_i-1|=0.086\pm0.078$, n=6, N=34; Figure~\ref{figure_3}O), corresponding to $D_{\textit{eff,t}}\approx0.71\pm0.06$ mm during the first few milliseconds (Figure \ref{figure_3}P). These post-launch dynamics are consistent with an impulsive release that launches adult droplets into an inertia-influenced deformation regime.

\subsection*{Droplet ejection by nymphs using a hinge-like stylus}
In nymphs, honeydew ejection consists of two phases, droplet growth followed by droplet ejection, without the metastable plateau or the loading phase observed in adults (Figure~\ref{figure_3}A-D; Supplementary Video 4). The split stylus opens during growth and closes to shed the droplet, behaving akin to a hinge about a pivot near its base (Figure~\ref{figure_3}B,E). A single ejection event lasts $\sim$15--20~ms, roughly an order of magnitude shorter than in adults (Figure~\ref{figure_3}F). 

Despite the shorter event duration, stylus kinematics remain ultrafast. Peak angular speed and acceleration reach $\dot\theta_\textit{max}\sim 10^5$~$^\circ$/s and $\ddot\theta_\textit{max} \sim 10^9$~$^\circ$/s$^2$, corresponding to tip accelerations on the order of  $L_\text{nymph}\ddot{\theta}_{\text{max}}\sim 15000$ m/s$^2$ ($\sim1500\,g$), about 50\% higher than in adults (Figure \ref{figure_3}G,H). Peak angle, peak angular speed, and droplet separation occur in sequence within $\sim$3~ms (Figure~\ref{figure_3}G), consistent with rapid hinge closure rather than delayed release from elastic energy storage.

Droplet launch, however, remains strongly subpropulsive. The stylus tip reaches $V_s= 1.27\pm0.26$~m/s while droplets launch at $V_d=0.22\pm0.01$~m/s, giving $\lambda=V_d/V_s=0.18\pm0.04$ (Figure~\ref{figure_3}M; $n=3$, $N=36$). Nymphs also eject in rapid succession with short closed and open intervals (Figure~\ref{figure_3}I,J), indicating efficient clearance without a prolonged recovery or elastic loading phase. 

Post-launch droplets show minimal deformation and remain nearly spherical in flight (Figure~\ref{figure_3}C,N; Table~\ref{table3};Supplementary Video~5). The normalized projected area changes are small (mean $|A_t/A_i-1|=0.06\pm0.05$; Figure~\ref{figure_3}O), about 25--30$\%$ lower than in adults, and droplets detach at $D_{\mathrm{eff,i}}\approx0.61\pm0.07$~mm (Figure~\ref{figure_3}P). 

Together, these measurements show that although nymphs achieve adult-like peak stylus kinematics, droplet launch is slower, more strongly subpropulsive, and produces minimal post-launch deformation, in contrast to the loaded, impulsive release observed in adults.

\begin{figure*}[t!]
\centering
\includegraphics[scale=0.6]{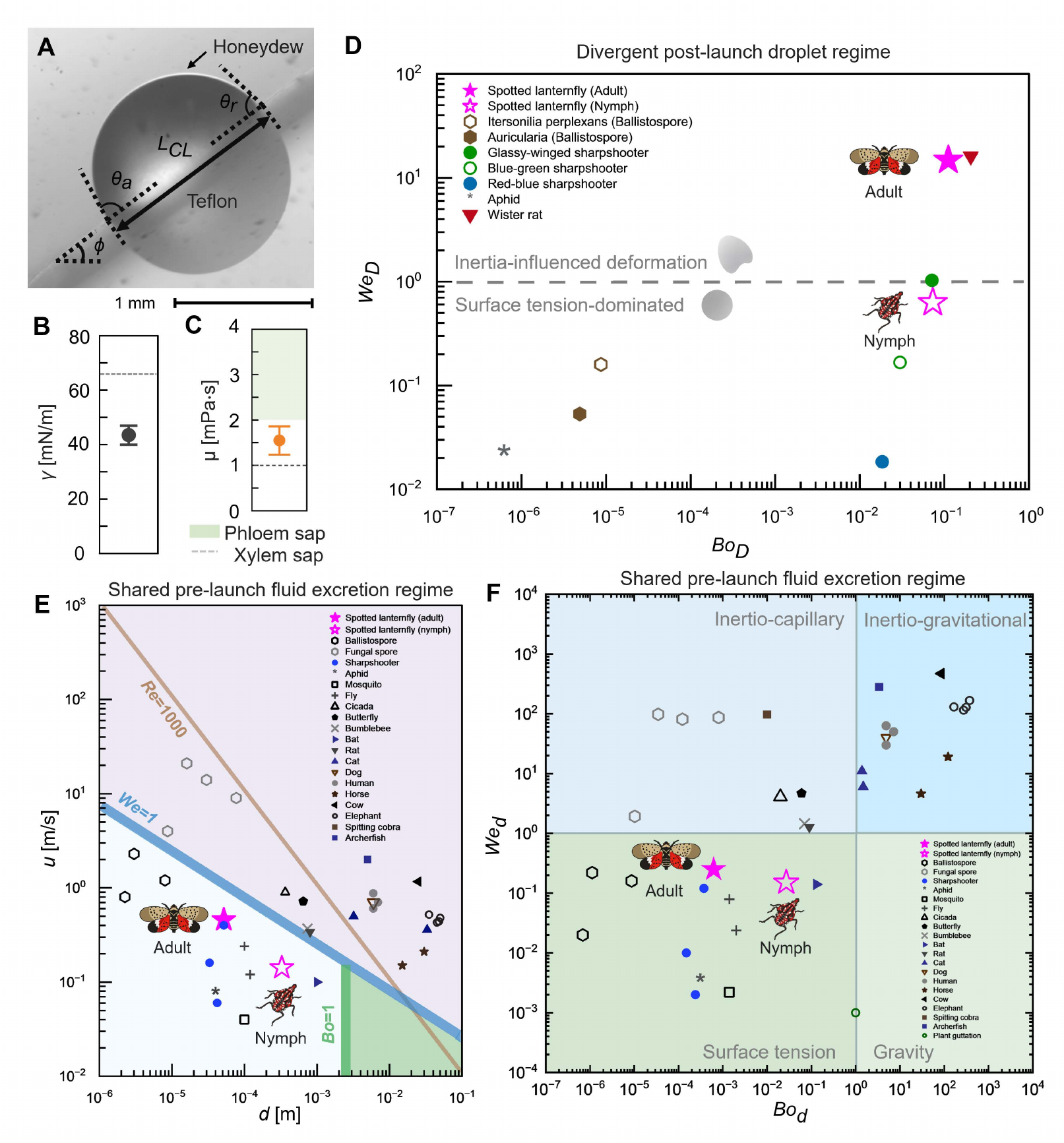}
\caption{\textbf{Measured honeydew properties place nymphal and adult SLF under similar pre-launch capillary constraints but in different post-launch droplet regimes.} (A) Tilted-plate method used to estimate the surface tension of SLF honeydew. A droplet on Teflon depins at a critical tilt angle $\phi$; measuring the contact-line length $L_\text{CL}$, and contact angle hysteresis $\cos\theta_r-\cos\theta_a$, provides a force balance to estimate $\gamma$ (see Mathematical Model). (B) Measured surface tension of SLF honeydew, $\gamma$ ($n=5$ droplets). (C) Measured dynamic viscosity $\mu$ of SLF honeydew ($n=6$ trials), with a gray dotted line indicating the viscosity of xylem sap and the light green area is the viscosity range of phloem sap. Error bars show mean $\pm one$ standard deviation. (D) Post-launch droplet regime map using droplet-based Bond and Weber numbers ($Bo_D$, $We_D$), computed from the initial effective droplet diameter $D$ and launch speed $V_d$. Adult droplets occupy an inertia-influenced deformation regime, whereas nymphal droplets remain surface-tension dominated with weak deformation. (E) Characteristic outlet speed $u$ and outlet diameter $d$ for nymphal and adult SLF, shown together with other fluid-ejecting organisms. Thick segments indicate the range spanned by representative Newtonian fluids due to differences in $\gamma$ and $\mu$. (F) Pre-launch Weber–Bond regime map showing that both nymphal and adult SLF lie in the same surface-tension-dominated regime ($We_d<1$, $Bo_d<1$), despite their different post-launch outcomes. Comparative organismal data in E and F are reproduced/adapted from refs~\cite{sharpshooter2023elio,Fluidejection2024elio}.}
\label{figure_4}
\end{figure*}

\begin{figure*}[t!]
\centering
\includegraphics[scale=0.55]{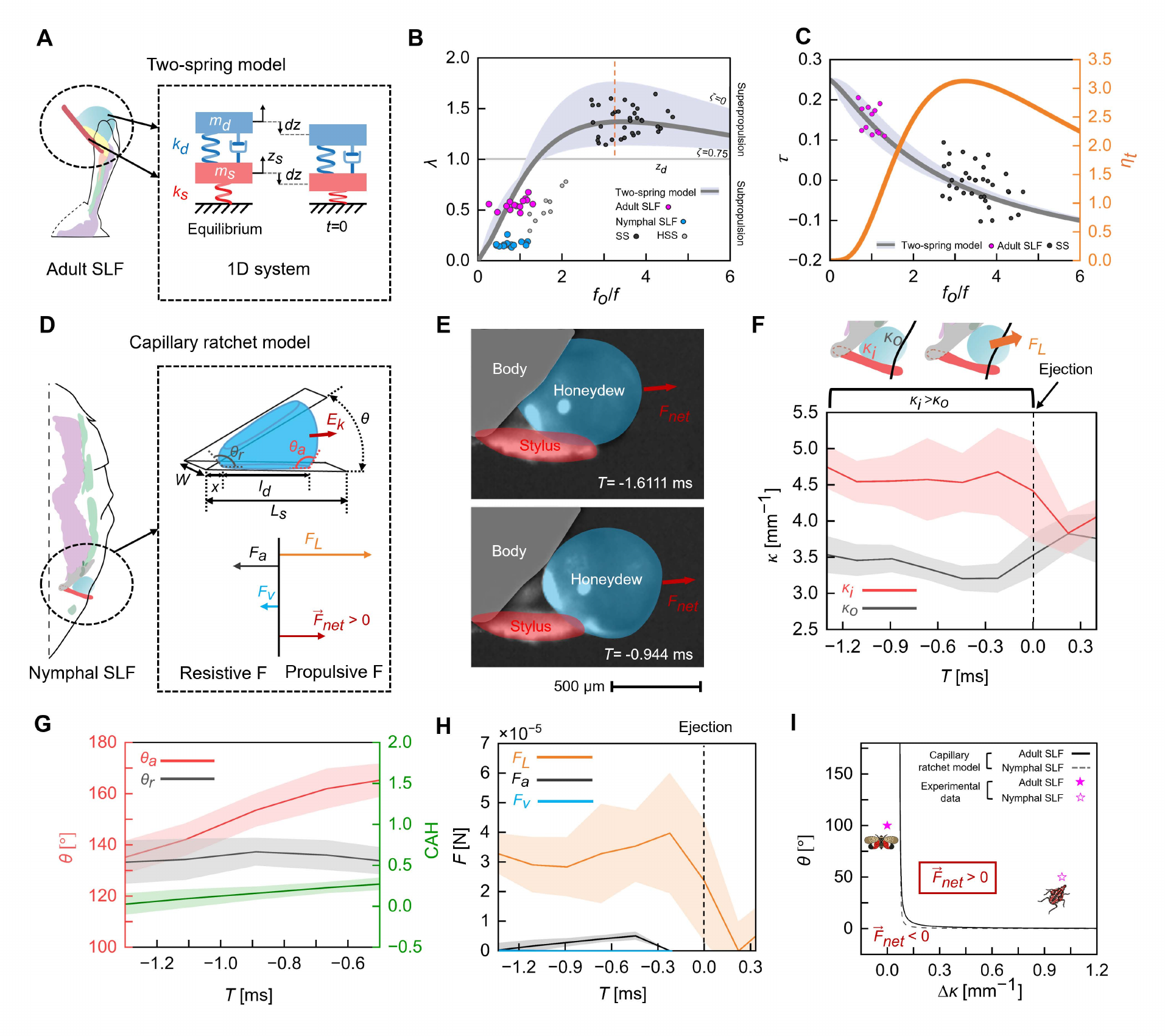}
\caption{\textbf{Adult SLF are described by a subpropulsive two-spring launcher, whereas nymphal SLF release droplets by capillary rectification.}  (A-C) Reduced-order two-spring model for adult SLF. Colored regions indicate the stylus, basal elastic element, muscle bundles, and gut channel. (A) Schematic of the coupled stylus-droplet oscillator, in which droplet ($m_d$,$k_d$) is treated as a capillary spring coupled to the stylus/actuator ($m_s$,$k_s$). (B) Speed ratio $\lambda=V_d/V_s$ plotted against the frequency ratio $f_o/f$, comparing xylem sap-feeding adult sharpshooters (SS) and hairless variant (HSS), with phloem sap-feeding adult and nymphal SLF. The orange dotted line corresponds to peak kinematics assuming no dissipation ($\zeta=0$) at $f_o/f\sim3.23$ and the solid gray curve denotes the theoretical curve for $\zeta=0.25$. 
(C) Dimensionless delay $\tau = (t_{c,\max} - t_{\dot{\theta},\max})/T_s$ and normalized energy transfer efficiency $\eta_t$ as functions of $f_o/f$ for the theoretical model, adult SLF, and SS, where $T_s=1/f$. 
(D) Schematic of the Laplace-pressure-driven droplet release by nymphal stylus during closure. (E) High-speed images used to extract interface curvatures and dynamic contact angles during detachment. (F) Temporal variations of inner and outer curvatures ($\kappa_i$ and $\kappa_o$). (G) Advancing and receding contact angles ($\theta_a$ and $\theta_r$) and dynamic contact angle hysteresis, $CAH_D=\cos\theta_r-\cos\theta_a$.  (H) Estimated force magnitudes along the stylus surface: Laplace driving force $F_L$, adhesion force $F_a$, and viscous resistance $F_v$ (nymphal SLF: $n=2$, $N=26$). Time is reported relative to droplet separation, with $T$=0 defined at the moment of ejection (dotted line). Solid lines show means and shaded regions indicate $\pm1$ SD. (I) Net-force map in the $\Delta\kappa$--$\theta$ plane for the capillary rectification model. Curves denote the $F_{\mathrm{net}}=0$ boundary for adult and nymph geometries; experimental points are overlaid for comparison.   
}
\label{figure_5}
\end{figure*}

\subsection*{Surface tension and viscosity of SLF honeydew}
Honeydew is a chemically complex secretion whose composition can vary with diet and physiology~\cite{SLF2022hajar}. To set material properties for scaling analysis and mathematical modeling, we measure the surface tension $\gamma$ and dynamic viscosity $\mu$ of SLF honeydew (Figure~\ref{figure_4}A). Since honeydew droplets from nymphs and adults cannot be reliably separated during collection, we report bulk properties and treat honeydew as a single Newtonian fluid.
 
We estimate surface tension using a tilted-plate method on Teflon, where the critical tilt angle and contact angle hysteresis provide a force balance for $\gamma$ (Figure \ref{figure_4}A,B; Mathematical Model section). Across five droplets, we obtain $\gamma=44$~mN/m on average. We measure dynamic viscosity with a micro-viscometer to obtain $\mu=1.55$~mPa$\cdot$s, on average across six trials (Figure \ref{figure_4}C), indicating that under our assay conditions fresh honeydew is only modestly more viscous than water.

\begin{figure*}[t!]
\centering
\includegraphics[scale=0.7]{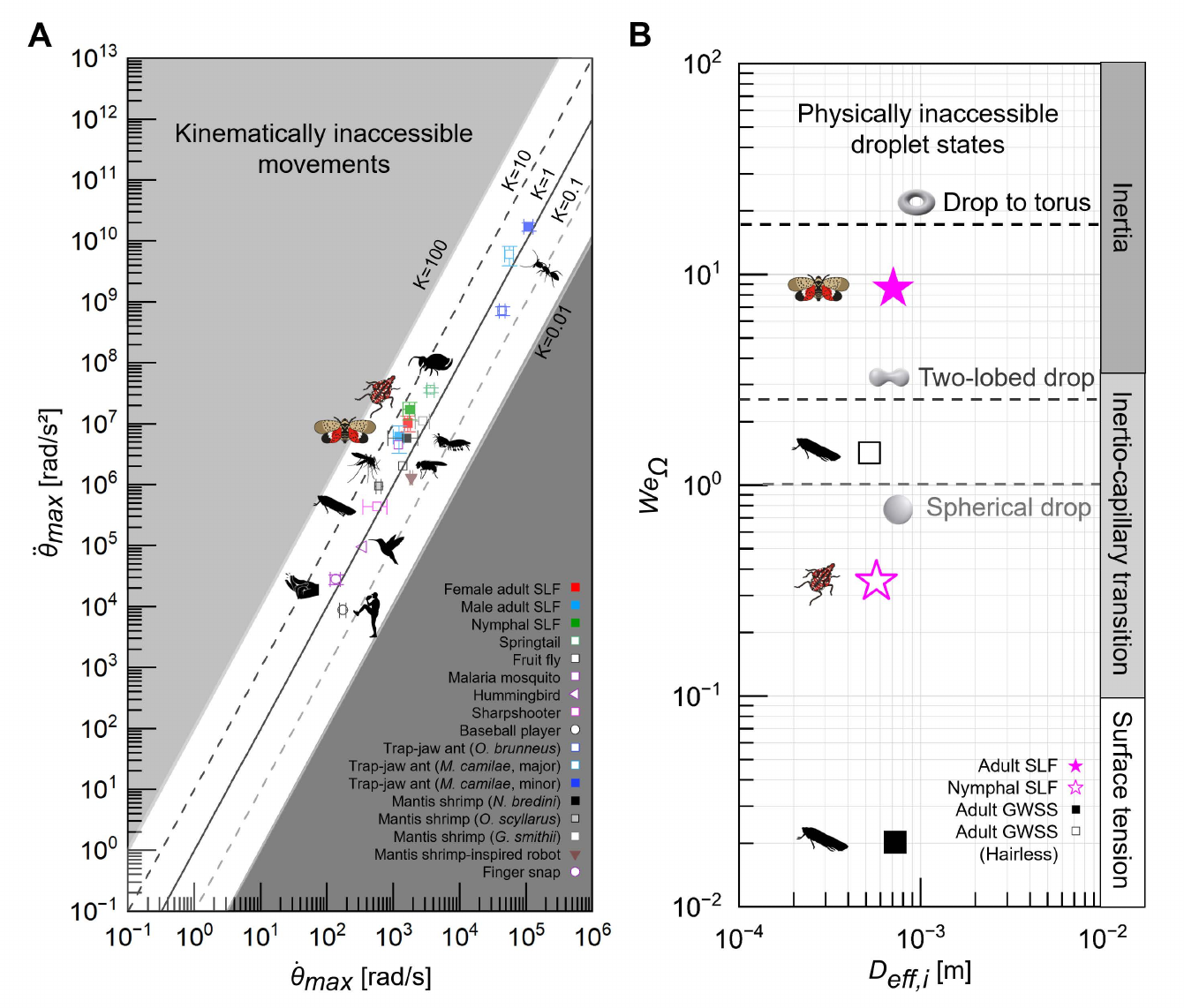}
\caption{\textbf{SLF stylus kinematics lie near the upper envelope of ultrafast biological rotation.} (A) Maximum angular speed ($\dot{\theta}_{\max}$) and maximum angular acceleration ($\ddot{\theta}_{\max}$) for ultrafast biological rotational systems~\cite{sharpshooter2023elio, springtail2022victor, springtail2024jacob, mosquitofruitfly2022wouter, fruitfly2014florian, mosquito2017florian,  hummingbird2018rivers, ballpitch1994glenn, ballpitch2006eric, mantis_shrimp_2021_emma,trap_jaw_ant_2018_fredrick,trap_jaw_ant_2022_gregory, finger_snap_2021_raghav}. Dashed lines denote contours of constant $K$ in $\ddot{\theta}_{\max}=K\dot{\theta}_{\max}^{2}$, and shaded regions indicate kinematically inaccessible combinations of speed and acceleration. Symbols show means and error bars denote $\pm1$ SD; source values are listed in Table~S5 (female adult SLF: $n=3$, $N=37$; male adult SLF: $n=3$, $N=33$; nymphal SLF: $n=3$, $N=36$). (B) Post-launch rotational regime map showing the spinning Weber number, $We_{\Omega}=\rho \Omega^{2}D^{3}/8\gamma$, as a function of initial effective droplet diameter $D$, for SLF and adult glassy-winged sharpshooters (GWSS). Dotted horizontal lines mark classical rotating-drop stability boundaries separating approximately spherical, two-lobed, and drop-to-torus states \cite{rotating_drop_stability_1965_subrahmanyan,rotating_drop_stability_1980_robert}. Adult SLF occupy a high-spin, inertia-influenced regime associated with nonaxisymmetric droplet deformation, whereas nymphal SLF remain in a surface-tension-dominated regime in which droplets remain nearly spherical after release.}
\label{figure_6}
\end{figure*}

\subsection*{Shared pre-launch capillary constraints and divergent post-launch droplet dynamics}
Droplet detachment separates honeydew ejection into two regimes. Before release, the droplet is an attached meniscus governed by outlet-scale capillarity and pinning and post-launch, it is a free body in flight, whose shape and motion are set by the competition between inertia and surface tension.
Using the measured honeydew $\gamma$ and $\mu$, we plotted nymphal and adult SLF within these pre-launch and post-launch regime maps for biological fluid ejection and compared them with other organisms (Figure \ref{figure_4}D--F). The pre-launch uses the outlet diameter $d$ and characteristic fluid exit speed $u$ to compute $Re_d=\rho u d/\mu$, $We_d=\rho u^2 d/\gamma$, and $Bo_d=\rho g d^2/\gamma$. Both nymphal and adult SLF fall in the same surface-tension-dominated pre-launch regime, satisfying $We_d<1$ and $Bo_d<1$. Adults have $We_d\approx0.25$ and $Bo_d\sim O(10^\text{-4})$, whereas nymphs have $We_d\approx0.15$ and $Bo_d\sim O(10^\text{-2})$. Thus, despite clear morphological differences, both life stages encounter similar capillary constraints before release.

The post-launch regime, however, is different for nymphs and adults (Figure~\ref{figure_4}D). Droplets ejected by adults deform strongly and rotate after launch (Figure~\ref{figure_2}C,L), whereas, nymphal droplets rotate but remain nearly spherical (Figure~\ref{figure_3}C,N). Adult droplets also carry substantially higher translational and rotational kinetic energies than nymphal droplets (Table~\ref{table3}). This separation is captured by the droplet-scale $We_D=\rho v^2 D/\gamma$, defined using the launch speed $V_d$ and initial effective droplet diameter $D$. Adults operate at $We_D$=14.8$>$1, where inertia competes strongly with surface tension and post-launch deformation is expected, whereas, nymphs operate at $We_D=0.63$, where surface-tension rapidly restores shape and deformation remains weak (Figure~\ref{figure_4}D). The developmental transition therefore does not reflect a change in the outlet-scale fluid regime. It reflects a change in how stylus motion is transferred into the droplet at release.

\subsection*{Adult SLF follows a subpropulsive two-spring catapult}
The adult kinematics suggest a loaded release rather than a purely kinematic flick. The droplet remains attached during rapid stylus motion, maximum droplet compression follows peak stylus speed by a finite delay, and the droplet leaves only after this compression phase. These features are the signatures expected from a coupled stylus--droplet oscillator, which motivates comparison with the reduced two-spring model developed for xylem-feeding sharpshooter excretion (Figure~\ref{figure_5}A--C)~\cite{sharpshooter2023elio}.

In this model, the droplet is treated as a capillary spring with natural frequency $f_o$, coupled to a stylus oscillator with frequency $f$. The key control parameter is the frequency ratio $f_o/f$, which sets the phase relationship between stylus motion and droplet compression and therefore the efficiency of energy transfer at release (see Methods for model details). Adult sharpshooters operate in the superpropulsive optimum, with $f_o/f\sim3.2$, where droplet speed exceeds stylus speed ($\lambda>1$) (Figure~\ref{figure_5}B). By contrast, both nymphal and adult SLF lie well below this optimum, with $f_o/f=0.77\pm0.23$ ($n=2, N=17$) for nymphs and $f_o/f=0.79\pm0.39$ ($n=4, N=14$) for adults. SLF therefore operate in a subpropulsive regime, placing them on the low-frequency space of the model. 

Adult SLF measurements collapse onto the theoretical two-spring response in this subpropulsive branch, whereas nymphal SLF deviate from it, similar to `hairless' sharpshooters that do not exhibit superpropulsion~\cite{sharpshooter2023elio}. The dimensionless delay between peak stylus speed and maximum droplet compression, $\tau=\frac{t_{c,\max}-t_{\dot{\theta},\max}}{T_s}$, where $T_s=1/f$ falls within the model range for adults, with $\tau=0.15\pm0.03$ ($n=3, N=12$) (Figure~\ref{figure_5}C). The corresponding normalized transfer efficiency ($\eta_t$), quantifying an energy transfer factor after take-off is substantially lower than for sharpshooters, consistent with the fact that adult SLF launch droplets subpropulsively rather than superpropulsively. These results indicate that adult SLF use a catapult-like mechanism that is well described by a coupled stylus--droplet oscillator, but in a regime optimized for droplet release rather than maximal speed amplification.

\subsection*{Nymphal SLF eject honeydew by capillary rectification}
Nymphs lack an obvious basal elastic element, their droplet--stylus contact time is short, and their measurements do not collapse onto the two-spring model prediction. We therefore model the nymphal stylus as a capillary rectifier. Here, the stylus forms an asymmetric wedge that biases capillary pressure during closure and rectifies liquid motion without requiring inertial launch (Figure~\ref{figure_5}D-I)\cite{capillaryratchet2008manu, 3D_capillary_ratchet2021shile}. As the stylus closes, the droplet acquires a larger inner curvature than outer curvature, $\kappa_i>\kappa_o$ (Figure \ref{figure_5}E,F), producing a Laplace pressure difference $\Delta P=\gamma(\kappa_i-\kappa_o)$ and a forward driving force along the stylus. High-speed measurements further show that both advancing and receding contact angles remain above $90^\circ$ during detachment, indicating a partially non-wetting regime (Figure.~\ref{figure_5}G). Dynamic contact angle hysteresis remains finite, but it stays below 0.5 immediately before detachment, so the corresponding adhesive force is smaller than the Laplace driving force.

Specifically, the Laplace driving force ($F_L=2\gamma\pi(\kappa_i-\kappa_o)D^2/4\sim$ O(10$^{-5}$) N) is an order of magnitude larger than the capillary adhesion force ($F_a=\gamma W (cos\theta_r-cos\theta_a)\sim$ O(10$^{-6}$) N) ($F_L>F_a$), and viscous forces are negligible $F_v=\mu Wl_d v/(\theta x)\sim$O$(10^{-8})$~N (Figure~\ref{figure_5}H). Using the measured geometrical, rheological, and kinematic parameters of the stylus-droplet system, we compute a net-force balance along the stylus and find that the nymphal ejection lies in a persistently positive regime relative to the theoretical $F_\text{net}=0$ boundary (Figure~\ref{figure_5}I). Thus, throughout detachment the net capillary force remains directed toward release. Nymphs therefore shed droplets by capillary rectification, whereas adults rely on elastic loading and release to launch droplets into an inertia-dominated post-launch regime.  

\section*{Discussion}
\vspace{-0.5\baselineskip} 
\subsection*{Summary} 
Across ontogeny, spotted lanternflies eject honeydew with two mechanically distinct release strategies. In nymphs, the stylus imposes an asymmetric capillary bias during closure and sheds droplets that remain surface-tension stabilized after takeoff. In adults, a longer stylus associated with an elastic basal region preserves droplet contact through a finite compression phase and launches droplets into a deformation-and spin-dominated regime. This interpretation is supported by terminal-abdomen morphology from micro-CT and SEM, high-speed measurements of stylus and droplet kinematics, direct measurements of honeydew surface tension and viscosity, and reduced-order mathematical models for adult and nymphal release. Both stages therefore solve the same fundamental outlet-scale capillary problem, but they do so with different modes of stylus--droplet coupling.

\begin{table*}[t!]
  \centering
    \caption{Parameters and estimates used in the energy-balance analysis of honeydew excretion in nymphal and adult SLF.}
    \footnotesize
      \resizebox{1.8\columnwidth}{!}{%
    \begin{tabular}{ccccc}
      \toprule
      \textbf{Parameter} & \textbf{Description} & \textbf{Nymphal SLF} & \textbf{Adult SLF} & \textbf{Unit} \\
      \midrule
      $\eta_p$ & Estimated Energy density of phloem sap & \multicolumn{2}{c}{$3.4\times 10^9$} & J/m$^3$ \\
      $\eta_f$ & Feeding cost density & \multicolumn{2}{c}{Negligible} & J/m$^3$ \\

      $\eta_e$ & Estimated  excretion cost density & $5.80\times 10^2$ & $1.26\times 10^4$ & J/m$^3$ \\
      $D_{\mathrm{eff,i}}$ & Initial effective droplet diameter & 0.61$\pm$0.07 & 0.71$\pm$0.06 & mm \\
      $\Delta t$ & Droplet growth time & 10.95$\pm$0.91  & 170.35$\pm$60.66 & ms \\
      $d$ & Distal outlet diameter & 0.34$\pm$0.05 & 0.05$\pm$0.01 & mm \\

      $l$ & Hindgut  length & \multicolumn{2}{c}{$\sim$1} & mm \\
      $\mu$ & Dynamic viscosity & \multicolumn{2}{c}{1.55} & mPa$\cdot$s \\
      $\gamma$ & Surface tension & \multicolumn{2}{c}{44} & mN/m \\

      $u$ & Flow speed in the hindgut & 0.14$\pm$0.04 & 0.46$\pm$0.11 & m/s \\
      \bottomrule
    \end{tabular}
    }
  \label{table5}
\end{table*}

\subsection*{Stylus rotation in SLF approaches the upper envelope of biological actuation}
How fast must, and how fast can, an excretory appendage rotate to detach and eject a sticky droplet at millimeter scale? To place the SLF stylus in context, we compare it with fast-rotating appendages used for strike, jump, escape, and rapid fluid handling (Figure~\ref{figure_6}A)~\cite{sharpshooter2023elio, springtail2022victor, springtail2024jacob, mosquitofruitfly2022wouter, fruitfly2014florian, mosquito2017florian,  hummingbird2018rivers, ballpitch1994glenn, ballpitch2006eric, mantis_shrimp_2021_emma,trap_jaw_ant_2018_fredrick,trap_jaw_ant_2022_gregory, finger_snap_2021_raghav}. In both nymphs and adults, peak angular speeds $\dot{\theta}_\text{max}$ are of order $10^3$~rad\,s$^{-1}$ and peak angular accelerations $\ddot{\theta}_\textit{max}$ of order $10^7$~rad\,s$^{-2}$. These values exceed those reported for the stylus of xylem-feeding sharpshooters and overlap the range occupied by globular springtail furca and mantis shrimp appendages~\cite{sharpshooter2023elio, springtail2024jacob, mantis_shrimp_2021_emma}. 

The SLF stylus lies in the narrow band of the $\dot{\theta}_{\max}$--$\ddot{\theta}_{\max}$ plane occupied by other ultrafast biological actuators, approximated by
\[
\ddot{\theta}_{\max}=K\dot{\theta}_{\max}^{2}
\]
where $K$ captures the effects of stroke amplitude, joint geometry, and mass distribution as a lumped parameter (see SI discussion). For a non-optimized, uniform, and homogeneous slender appendage, one expects $K$ on the order of $10^{-2}$ to $10^{-1}$. However, hollowing and proximal mass concentration move biological systems upward into the empirically occupied band. Both nymphal and adult SLF fall within that higher-$K$ regime, with $1<K<10$, approaching the putative ceiling imposed by elastic-wave propagation and mechanical coherence~\cite{physicallimit_2005_jan}. Both nymphal and adult SLF thus already operate in the regime of highly optimized biological actuators. The developmental difference must therefore lie less in how fast the stylus can move than in how that motion is coupled to the droplet and converted into release.

\subsection*{Rotational Weber number separates adult and nymphal droplet states}
The mechanical relevance of the SLF stylus timescale is clearer when referenced in context to the honeydew droplet dynamics. For droplets of the size ejected by SLF, the inertio-capillary relaxation timescale is only a few milliseconds, comparable to the interval over which stylus acceleration, droplet compression, and release occur. Stylus--droplet coupling therefore unfolds under strongly dynamic wetting and dewetting conditions, for which viscous and contact-line dissipation are unlikely to be negligible.

The post-launch regime is better distinguished by droplet spin than by stylus kinematics alone. To examine this regime, we use the spinning Weber number, $We_\Omega=\rho \Omega^2 D_{\mathrm{eff},i}^3/8\gamma$, where $\Omega$ is the angular velocity of the droplet after ejection and $D_{\mathrm{eff},i}$ is the initial effective droplet diameter (Figure~\ref{figure_6}B, Table~\ref{table3}). In contrast to the translational Weber number, which describes the balance between inertia and capillarity during launch, $We_\Omega$ asks whether the droplet can remain shape-stable once rotational inertia has been established. Nymphal droplets fall in the range $We_\Omega<$1, where capillary restoring forces exceed centrifugal stresses, and accordingly retain a nearly spherical shape throughout despite their motion. Adult SLF droplets lie above that boundary, with $We_\Omega>$1, entering the regime in which rotational inertia is strong enough for nonaxisymmetric deformation to compete with capillary restoration~\cite{rotating_drop_stability_1965_subrahmanyan, rotating_drop_stability_1980_robert}. The elongated, two-lobed, peanut-like morphologies seen in our high-speed sequences are consistent with that placement (Figure~\ref{figure_2}C). Relative to glassy-winged sharpshooters, adult SLF occupy a higher-spin droplet regime. This separation likely reflects differences in droplet size and ejection mechanics associated with the broader distinction between xylem and phloem feeding.

\subsection*{Excretion costs are small relative to the energy density of the phloem meal}
Using the measured droplet size, growth time, outlet diameter, viscosity, and surface tension to estimate hindgut flow speed and the associated viscous and capillary pressure costs, we obtain an excretion cost density of $\eta_e \simeq 1.26\times10^{4}$~J\,m$^{-3}$ for adults and $\eta_e \simeq 5.80\times10^{2}$~J\,m$^{-3}$ for nymphs (Table~\ref{table5}). These values remain far below the estimated chemical energy density of the phloem meal, $\eta_p \simeq 3.4\times10^{9}$~J\,m$^{-3}$. At the scale of a feeding--excretion cycle, the energetic burden of honeydew removal is therefore small relative to the energy content of the incoming fluid. That contrasts with xylem-feeding sharpshooters, where the nutritional return of the meal is much lower, since xylem sap is overwhelmingly water (95\%) and must be acquired under negative pressure~\cite{phloemxylemcompare1983Elisabeth,sharpshooter2023elio}; under those conditions, improvements in ejection efficiency such as resonance-tuned superpropulsion carry greater energetic consequences~\cite{sharpshooter2023elio}. The much higher energy density of phloem likely relaxes the selective pressure for maximal ejection efficiency, which may help explain why both nymphal and adult SLF remain subpropulsive.

We note that these estimates are an order-of-magnitude excretory cost and not a detailed metabolic budget. For example, they do not include the energetic demands associated with osmotic regulation, probing and selecting feeding sites, salivary secretion into host tissue, grooming after excretion, or maintenance of nutritional symbionts~\cite{problemsofphloemfeeding2006angela, osmoticpotential2011julien, phloemsearching2021vamsi, alterprobingpattern2016yucheng, lanternbug2007piotr, grooming2013marianna, plantdefense2013torsten}. They also do not account for xylem drinking used by some phloem-feeding hemipterans to dilute the meal during osmoregulation~\cite{problemsofphloemfeeding2006angela, osmoticpotential2011julien}. Accessing a mixed diet across multiple hosts increases survival and fecundity~\cite{SLFannualsreview_2023_urban}, suggesting that intermittent xylem feeding may be critical for preventing excretory failure under heavy feeding loads. Those terms would need to be included in a full organismal budget. Nonetheless, for the comparison developed here, the main point remains unchanged: relative to xylem-feeding sharpshooters, SLF clear waste in a much less energetically constrained setting, and a subpropulsive operating point is therefore not difficult to reconcile with the available energetic margin.

\subsection*{Stylus-droplet coupling changes with ontogeny}
The capillary constraint at the outlet does not change across ontogeny: in both nymphs and adults, droplet formation and detachment occur at $We_d<1$ and $Bo_d<1$. The divergence emerges only when the droplet is about to separate. 

In nymphs, the direction of release is set by capillary geometry. During closure, the stylus forms an asymmetric wedge that maintains a curvature difference between the inner and outer interfaces, generating a Laplace-pressure bias toward ejection, while dynamic contact-angle hysteresis suppresses reverse motion during detachment. The resulting force remains directed along the stylus, so hinge closure alone is sufficient to produce directional transport. The stylus thus acts as a `geometric rectifier', converting a symmetric hinge motion into a net directional flux of liquid. This mechanism is analogous to capillary ratchets based on asymmetric confinement and wetting hysteresis (Figure~S3), including the directional transport described for shorebird beaks~\cite{capillaryratchet2008manu,3D_capillary_ratchet2021shile}, although it does not appear to rely on repeated depinning cycles.

The energy-storage structure for the rapid nymphal closure itself remains less well resolved. The available morphology and kinematics argue against an adult-like resilin-based launcher, but do not exclude smaller or more distributed elastic or muscle contributions within the hinge cuticle. However, what the present data establish is that capillary geometry and wetting asymmetry are sufficient to account for the observed directionality of release.

Why adults depart from the nymphal mode remains unresolved. As body size increases, both sap throughput and the volume of honeydew that must be cleared increase. Relative to nymphs, adults transfer a larger fraction of stylus speed into the droplet (Figs.~\ref{figure_2}I,J and \ref{figure_3}M), maintain stylus--droplet contact through a finite compression phase (Fig.~\ref{figure_2}B,G), and release a droplet in an inertia-influenced translational and rotational regime (Figs.~\ref{figure_4}D and \ref{figure_6}B). Comparable ontogenetic changes in spring-actuated systems have been described in juvenile snapping shrimp, where the elastic snapping mechanism emerges during post-settlement development and juveniles are already capable of cavitation at millimeter scale~\cite{harrison2023developing}. In SLF, the capillary problem at the outlet persists across ontogeny, but the demands placed on release after detachment do not. One likely consequence is a lower tendency for honeydew to wet the terminal abdomen and nearby plant surface, with corresponding effects on sooty mold growth and the local chemical environment around feeding aggregations~\cite{lanternbug2007piotr,de2024honeydew,SLFlure2024miriam}. Whether a similar developmental redistribution of function occurs elsewhere within Fulgoridae, or is more specific to \textit{Lycorma}, remains an open question. This morphological shift may represent a gradual ontogenetic program across the nymphal instars, culminating in a distinct binary switch to the elastic catapult in adulthood. Furthermore, the presence of structurally similar anal stylus in nymphs of other fulgorids, such as \textit{Zanna tenebrosa}~\cite{Fulgorideaevolution_2009_emeljanov}, suggests that this capillary rectification mechanism may be broadly conserved among early-stage phloem feeders.

The adult mechanism also appears to operate within a finite droplet-size window. During high-speed imaging, adults occasionally failed to eject droplets once they became sufficiently large, and the retained droplets were then removed secondarily with the legs. A lower bound is expected for the opposite reason, since capillary adhesion should increasingly impede clean detachment as droplet size decreases. The similar droplet sizes used by nymphs and adults despite different outlet diameters (Fig.~\ref{figure_6}B; Table~\ref{table3}) are at least consistent with such a window, although its boundaries remain to be mapped explicitly. Host plant, feeding state, and temperature could shift those bounds through their effects on honeydew composition and rheology \cite{SLFannualsreview_2023_urban}. As first through third instars feed on tender tissues of a broad range of hosts while fourth instars and adults narrow on specific woody tissues~\cite{SLFannualsreview_2023_urban}, honeydew properties in early nymphs may differ from the later stages. Because honeydew properties were measured here in bulk rather than by life stage or host plant, the present analysis likely understates how much chemical variation could alter the mechanics of release.

\subsection*{Concluding remarks}
Spotted lanternfly is now established across much of the eastern United States, spreads partly through human-mediated dispersal, and threatens fruit and forest industries through broad host use and heavy phloem feeding~\cite{SLFAPHIS2025,ladin2023human}. Across infested landscapes, copious honeydew deposition is one of the most visible consequences of SLF feeding. On host plants and nearby surfaces, it promotes mold and contributes volatile cues that act with body-derived odors during aggregation, mate finding, and oviposition~\cite{SLF2015surendra,lanternbug2007piotr,honeydewchemicals2022hajar,SLFlure2024miriam,de2024honeydew}. The massive honeydew output likely amplifies these chemical signals, creating a positive feedback loop that facilitates mating success~\cite{SLFannualsreview_2023_urban}. The ejection mechanics resolved here therefore bear directly on where honeydew is deposited, how long it remains near the body and host surface, and how chemical cues and microbial growth are distributed around feeding aggregations. Developmental stage-specific release mechanics may prove relevant not only to SLF biology and management, including lure-based monitoring and trapping strategies, but also to the broader ecology of sugar-rich residues in infested systems~\cite{SLFlure2024miriam,de2024honeydew}. More broadly, the transition between capillary rectification and elastic launch identifies two design principles for handling contaminating liquids at small scales, with possible relevance to bioinspired droplet ejectors, self-cleaning surfaces, and microfluidic waste-clearing systems~\cite{capillaryratchet2008manu,3D_capillary_ratchet2021shile,sharpshooter2023elio}.

\section*{Data, Materials, and Software Availability}
\vspace{-0.5\baselineskip}
The experimental data and simulation results that support the findings of this study are available in the GitHub repository at https://github.com/bhamlalab/spotted-lantern-fly-2026. Source data for all figures are provided with this paper.

\section*{Acknowledgments}
\vspace{-0.5\baselineskip}
We thank Kelly Murman for her assistance with honeydew collection in the field. We thank Isaiah Canlas, Sam Stella, and Corrine Losch for their assistance with spotted lanternfly experiments at the Forest Pest Methods Laboratory. N.H. thanks Sungho Park for the discussion. S.B. acknowledges the funding support from NSF CAREER iOS-1941933 and the California Department of Food and Agriculture Project (25-0371). This material was made possible, in part, by a collaboration with the United States Department of Agriculture’s Animal and Plant Health Inspection Service (APHIS). It may not necessarily express APHIS’ views. This research used resources of the Advanced Light Source, which is a DOE Office of Science User Facility under contract no. DE-AC02-05CH11231.

\section*{Author Contributions}
\vspace{-0.5\baselineskip}
N.H., E.J.C., and J.S.H. collected high-speed imaging data and N.H. and K.E.L. analyzed the data; N.H. collected microscopic images.; E.J.C., M.C., and S.B. provided insect and field photographs.; N.H. performed the mathematical modeling based on the precedent work by E.J.C.; N.H. and E.G.C. collected micro-CT scans and N.H. analyzed them.; M.C. collected honeydew and N.H. analyzed its properties.; M.C. provided biological samples and facilities.; M.C. and S.B. initiated the project.; N.H. and S.B. wrote and edited the manuscript.; N.H., E.J.C., J.S.H., E.G.C., K.E.L., M.C., and S.B. reviewed the paper.; S.B. supervised funding resources.

\section*{Author Declarations}
\vspace{-0.5\baselineskip}
The findings and conclusions in this publication are those of the author(s) and should not be construed to represent any official USDA or U.S. Government determination or policy.

\section*{Additional Information}
\vspace{-0.5\baselineskip}
Supplementary Movies are available for this paper. Correspondence and requests for materials should be addressed to Saad Bhamla.

\FloatBarrier

\printbibliography[title={References}]
\end{refsection} 

\clearpage

\onecolumn
\section*{Supporting Information}

\begin{refsection}

\section*{Materials and Methods}
\vspace{-0.5\baselineskip}
\subsection*{Honeydew Preparation and Characterization}
Fresh honeydew was collected beneath an \textit{Ailanthus} tree in Stroudsburg, Pennsylvania, between 1:00 pm and 2:00 pm on 21 August 2023, using aluminum foil and a pipette (Figure S4). The honeydew was stored in wax-sealed capillary tubes to prevent evaporation. The dynamic viscosity ($\mu$) of the honeydew was measured using a viscometer (Rheosense microVISC-m). 
Before starting the measurements, we cleaned the viscometer microchannel using 1 $\%$ Aquet cleaning solution. Before measuring the honeydew samples, we first measured the viscosity of deionized water to verify the reliability of the setup. We obtained a value of 1.044$\pm$0.015 mPa$\cdot$s (mean$\pm$SD, n=15), corresponding to a 4.4 $\%$ deviation from the ideal value (1 mPa$\cdot$s). For the honeydew viscosity measurements, we used a pumping rate of 5000 s$^{-1}$ and a sample volume of 21 to 27 $\mu$L for six trials (Table~S3). For surface tension measurements, honeydew was collected using a pipette from plant chambers containing spotted lanternflies at the USDA Forest Pest Methods Laboratory in Buzzards Bay, MA [USDA permit \#526-23-107-88901]. The contact angle hysteresis (CAH) of honeydew was measured by placing a honeydew droplet on a tilted glass slide covered with Teflon tape and capturing side-view images of the droplet using a Chronos camera (Figure S5). The measured CAH and the contact-line lengths were then converted to surface tension values using the force-balance equation (Table~S2).

\subsection*{High-speed Imaging and Tomography}
We conducted observation experiments with a high-speed camera (FASTCAM MINI AX) with a macro lens (Canon MPE 65 mm) and a field portable Zaila high-intensity light at the USDA Forest Pest Methods Laboratory in Buzzards Bay, MA (Figure S5). We obtained datasets of honeydew excretion for nymphal SLF (4th instar, n=3, N=36) and adult (n=10 individuals, N=84 ejection) in controlled plant chambers. Both systems' resulting trajectory and displacement was assumed to be in 2D. 3D morphological structures of the hindgut of spotted lanternflies were observed using X-ray microcomputed tomography ($\mu$-CT) at beamline 8.3.2 (Tomography) of Advanced Light Source of Lawrence Berkeley National Laboratory. 4th instar nymphs (n=2) and adults (2 females, 2 males) of SLF preserved in 99$\%$ ethanol for $\mu$-CT imaging were obtained from the USDA Forest Methods Laboratory. Tomographic structures were reconstructed using Dragonfly software with the aid of deep learning-based segmentation tools within the software. Other microscopic images of spotted lanternflies were captured using the Phenom ProX Scanning Electron Microscope (SEM). The test samples were coated with platinum for 60 s to avoid the charge effect prior to SEM imaging.

\subsection*{Data analysis}
We used ImageJ software (FIJI) to process the high-speed image data. We manually tracked the angular displacement of the stylus with respect to the axis along the body of spotted lanternflies (SLF). The pivot point where the stylus rotates is determined as a location where three lines at different stylus angles intersect each other. To verify the rationality of this pivot point, we calculated the stylus length as the distance between the tip of the stylus and the pivot point and identified this value as similar to the value of the stylus length from micro-CT data (Figure S4). Although this pivot point is stationary for the nymph SLF through the entire droplet ejection process, it was not stationary for adult SLF when the stylus ejects the droplet. To offset the kinematics of this relative movement, we subtract the difference in displacement of the body from the initial pivot point. The maximum linear speed of the stylus is calculated as $V_s$=$L_s$$\dot{\theta}_\textit{max}$. Similarly, we manually track the centroid of the droplet after ejection defined as the moment when the droplet is completely detached from the stylus. We measure the 2D location of the droplet P($x_m$,$y_m$) over time. The displacement of the droplet is calculated by accumulating the Euclidean distance $d(t)=\sum_{t=t_e}^{t}\sqrt{(x_m(t)-x_e)^2 + (y_m(t)-y_e)^2}$ where $(x_e, y_e)$ is the centroid of the droplet at take-off time $t_e$. The droplet velocity for adult SLF is calculated as $V_d=d(t_f)/(t_f-t_e)$ where $t_f$ is the final time in the field of view. For nymphal SLF, the droplet speed is simply calculated as a linear slope of the droplet displacement plot over time. To extract the stylus frequency value, we examine the shape of the angular displacement curve. We approximate the stylus displacement as a step function and the data are fitted using the MATLAB curve fitting tool to the following kinematics model of $\theta(t)= a \times erf(b \times t +c)+d$ where a, b, c, and d are fitting factors. 
The frequency of stylus was extracted from the kinematics data by considering the peak-to-peak duration in the angular acceleration curve. 

\subsection*{Derivation of surface tension of honeydew}
To derive the surface tension of honeydew, we consider a droplet with spherical geometry pinned on the tilted surface under the gravitational field \cite{fogmesh2013kyoochul,foghydrogel2019nami}. For a honeydew droplet on the tilted Teflon surface, the pinning force caused by contact angle hysteresis, $F_\textit{CAH}$, is balanced with the gravitational force term along the tilted surface, $F_\textit{g,s}$, as follows:

{\footnotesize
\begin{equation}\label{Fc}
\begin{aligned}
F_\textit{CAH} &= (\textit{contact line})\cdot\gamma \cdot CAH = 2a \gamma (\cos \theta_r - \cos \theta_a)\\ 
F_\textit{g,s} &= (\textit{droplet volume})\cdot\rho_L g \sin\phi = \left(\frac{3}{4}\pi r_\textit{drop}^3\left(\frac{\theta}{\pi}\right)-\frac{\pi}{3}a^2h\right)\rho_L g \sin\phi
\end{aligned}
\end{equation}
}

\noindent where $\phi$ is the tilting angle, which indicates how much the surface under the droplet is tilted. $\gamma$ is the surface tension of honeydew droplets ejected from spotted lanternflies. Here, $\theta_a$ and $\theta_r$ correspond to the advancing and receding contact angles of the static droplet, respectively. The density of honeydew, $\rho_L$, is assumed to be the same as the water density, $\rho_L=\rho_w=10^3$ kg/m$^3$. Here $\alpha$ is the half of the contact lines as $\alpha$ = $L_\textit{CL}$/2. Pythagoras theorem satisfies the statement of $r_\textit{drop}=\alpha$/sin$\theta$ and $h$ = $\alpha/tan\theta$. By balancing the two force terms, we obtain the surface tension values as follows:
{\begin{align}
\gamma =\frac{\rho_wg\pi \sin\phi \alpha^2}{6(\cos\theta_r-\cos\theta_a)}\left(\frac{4\theta}{\pi \sin^3\theta}-\frac{1}{\tan\theta}\right)\:\:[mN/m]
\label{surface tension}
\end{align}}

where $\alpha$ = $L_\text{$CL$}$/2 and $\theta$ = ($\theta_r$+$\theta_a$)/2.

\subsection*{Dimensionless numbers for fluid ejection of Newtonian fluid}
For the pre-launch excretion regime, the following dimensionless numbers are essential to characterize the fluid ejection regime in air, where $u$ is the 1D fluid speed at the nozzle exit, $d$ is the diameter of the nozzle, $\rho$ is the fluid density, $g$ is the gravitational acceleration, $\gamma$ is the surface tension, and $\mu$ is the dynamic viscosity of fluid:
\begin{equation}
\begin{aligned}
    Bo_d &= \frac{\text{gravity}}{\text{surface tension}} = \frac{\rho g d^2}{\gamma} \\
    We_d &= \frac{\text{inertia}}{\text{surface tension}} = \frac{\rho u^2 d}{\gamma} \\
    Re_d &= \frac{\text{inertia}}{\text{viscous force}} = \frac{\rho u d}{\mu}
\end{aligned}
\end{equation}

For organisms emitting liquid waste in the type of water, urine, and honeydew, the viscosity and surface tension values among them satisfy $\gamma_\text{honeydew} \leq \gamma_\text{urine} \leq \gamma_\text{water}$ and $\mu_\text{water} \leq \mu_\text{urine} \leq \mu_\text{honeydew}$. Assuming that the density of the fluid is the same as the density of the water, the critical diameter of the nozzle and the fluid speed for each fluid to be $Bo$=1 satisfy $d_\text{honeydew} \leq d_\text{urine} \leq d_\text{water}$. Under the same density and diameter conditions of the nozzle, the critical fluid speed for each fluid to be $We$ = 1 satisfies $v \propto \sqrt{\gamma_\text{fluid}}$, hence $v_\text{honeydew} \leq v_\text{urine} \leq v_\text{water}$. Hence, dimensionless numbers of urine always exist between water and honeydew for $Bo_d$, $We_d$ and $Re_d$. 

For the post-launch regime, the Weber number is defined using a droplet diameter, $D$, and a droplet ejection velocity, $V_d$, as the characteristic length and velocity scales as $We_D$=(inertia/surface tension)= $\rho V_d^2 D/\gamma$. As a representative $D$ value, we use the initial effective diameter of the droplet, so $D=D_\mathrm{eff,i}$. It describes that inertial force dominates over surface tension once the droplet is in the air when $We_D>1$. Hence, the droplet can be stretched into an ellipsoid or flatten as it moves through the air. With enough inertia, the droplet can travel outward rather than fall due to gravity. On the other hand, for $We_D<1$, surface tension dominates over inertial force for the honeydew droplet in flight. In addition to the translational $We$ numbers, the spinning Weber number defined as $We_\textit{$\Omega$}=\rho \Omega^2 D^3/(8\gamma)$ informs the physical limits in movements of rotating droplets.

\subsection*{Energetics of honeydew excretion of phloem sap-feeding insect}
To understand the excretion energetics through one cycle of feeding sap and excreting honeydew, we describe the volumetric energy density ($\eta$) that SLF gains as $\eta\geq (\eta_p - \eta_f)-\eta_e \:\:[J/m^3]$ where $\eta_p$, $\eta_f$, and $\eta_e$ indicate the energy content of nutrients in the phloem sap, the energy per unit volume expended by the cibarial pump during pumping and the energy density required during honeydew excretion. For SLF, $\eta_f$ would not be as high as xylem feeding insects that need to suck xylem sap by actively generating negative pressure, since the phloem sap is under positive pressure, allowing insects to acquire nutrients with minimal muscular suction efforts \cite{phloemxylemcompare1983Elisabeth}. As the term $\eta_f$ is negligible in the energetic excretion relation for phloem feeders \cite{negligiblefeedingenergy_2006_torsten}, we can finally write the equation above as $\eta\geq \eta_p -\eta_e \:\:[J/m^3]$. Generally, the total solute concentration of phloem sap is approximately 20 wt.$\%$ and sugar constitutes $80\sim90\space \%$ of the solute \cite{phloemsugarconcent2013kaare}. At this concentration, the density of sugar aqueous solution, $\rho_s$, is 1.08 [g/ml] since the sugar mainly consists of sucrose \cite{phloemsapchemcomp1972Shelagh}, which is nearly the same with water density. Altogether, the sugar concentration, $C$, is estimated as $2.16\times10^5$ [g/m$^3$]. The specific energy released by chemical reaction of sugar, $E_s$, is 17,000 [J/g]. By multiplying the $E_s$ with the $C$ and $\rho_s$, we can derive the volumetric energy density of phloem sap ($\eta_p$) as 3.4$\times 10^9$ [J/m$^3$], much higher than the xylem sap energy density of $\eta_x \sim 10^5 - 10^6 \:\:$ [J/m$^3$] \cite{sharpshooter2023elio, xylemsapenergy2021Elisabeth}. To estimate the excretion energy term, $\eta_e$, the excretion of fluid through the anal tube can be modeled as a pressure-driven flow of honeydew through a straight and circular cylinder having an orifice diameter $d$ and anal length $l$ estimated from micro-CT scans \cite{sharpshooter2023elio, urine2014patricia}. Using the energy balance equation to pump water at a steady state with a flow speed $u$ across the hindgut, the energetic cost of pumping fluid is described as:   

{\begin{align}
\eta_e= \left(\frac{32\mu l u}{d^2}+\frac{4\gamma}{d}\right) 
\label{Eta_excretion}
\end{align}}

where $\mu$ and $\gamma$ represent the dynamic viscosity and surface tension of honeydew. The orifice diameter is expressed as $d=4A/P$ where $A$ and $P$ indicate the respective cross-sectional area and perimeter of the rectum,  obtained from micro-CT scans and SEM images \cite{sharpshooter2023elio, urine2014patricia} (Figure S1). The flow speed $u$ is expressed as: 

{\footnotesize\begin{align}
u\:= \frac{Q}{A}\:= \frac{V/\Delta t}{\pi d^2/4}\:= \frac{V}{\Delta t \pi d^2/4}\:=\frac{\pi D^3/6}{\Delta t \pi d^2/4}\:=\frac{2D^3}{3 d^2 \Delta t}
\label{u_flowspeedcylinder}
\end{align}}

where $Q$, $A$, and $\Delta t$ indicate the volumetric flow rate, cross-sectional area of the cylinder, and droplet growth time \cite{sharpshooter2023elio}. Here, the final droplet volume, $V$, is estimated as $V=\pi D^3/6$ where $D$ is the initial effective diameter of honeydew droplet upon ejection. Next, we explore the kinetic energy consumption of honeydew after droplet ejection. First, we describe the definition of the kinematic properties of the stylus and droplet. Since the honeydew droplet has an effective diameter, $D$, the droplet mass, $m_d$, is estimated to be $m_d \simeq \rho_w \pi D^3/6$. The angular velocity of rotating honeydew droplets, $\Omega$, can be roughly estimated as $\Omega\sim\Delta\theta/\Delta t\sim (\theta_2-\theta_1)/(t_2-t_1)$ where $t_1$ and $t_2$ are the time scales right after the ejection. The droplet rotation was assumed as 2-D rotation. Micro-particles that originally exist in the honeydew droplet were used to track the angular variation of the droplet within a short period of time. We used 0.17 ms and 0.6 ms for the respective nymphal and adult SLF to derive the estimation of the angular velocity. For honeydew droplets, the rotational kinetic energy, $E_R$, is defined as $E_R=I\Omega^2/2$ where $I$ is the moment of inertia. Using the moment of inertia of the droplets, $I\simeq km_dD^2$ where $k$ is a geometry-dependent constant, we can express the rotational kinetic energy as $E_R$=$I\Omega^2/2\sim k m_d D^2 \Omega^2/2$. For simplicity, we used 0.1 as the $k$ value, which is the case for a rigid sphere. The translational kinetic energy of the droplet, $E_K$, can be written as $E_K=m_d V_d^2/2\sim \rho_w \pi D^3 V_d^2/12$.

\subsection*{Two-spring model for adult SLF}
To interpret adult honeydew ejection within the same framework used for sharpshooter ejection~\cite{sharpshooter2023elio}, we model the adult stylus--droplet system as two coupled oscillators in 1D. The lower oscillator represents the stylus together with the basal elastic element, with mass $m_s$ and stiffness $k_s$. The upper oscillator represents the attached droplet, with effective mass $m_d$, capillary stiffness $k_d$, and damping $c_d$.

The coupled equations of motion in the vertical direction are:

{\begin{align}
M\ddot{z} + C\dot{z} + Kz = 0
\end{align}}
with 
\begin{equation}
\resizebox{0.5\linewidth}{!}{%
    $\displaystyle 
    \begin{bmatrix}
    m_s & 0 \\
    0 & m_d
    \end{bmatrix}
    \begin{pmatrix}
    \ddot{z}_s \\
    \ddot{z}_d
    \end{pmatrix}
    +
    \begin{bmatrix}
    c_d & -c_d \\
    -c_d & c_d
    \end{bmatrix}
    \begin{pmatrix}
    \dot{z}_s \\
    \dot{z}_d
    \end{pmatrix}
    +
    \begin{bmatrix}
    k_s + k_d & -k_d \\
    -k_d & k_d
    \end{bmatrix}
    \begin{pmatrix}
    z_s \\
    z_d
    \end{pmatrix}
    = 0
    $
}
\label{eq:two_spring} 
\end{equation}

Simulations begin with the  lower spring compressed by a displacement $dz_s$ below the equilibrium, representing the deformation of the resilin-based basal elastic element during the spring loading phase. At $t=0$, the  lower spring is released and the coupled dynamics are integrated  numerically  using a fourth-order Runge--Kutta method in MATLAB. The droplet  is modeled as an effective  Hookean spring. Its spring constant is is $k_d=\frac{32\pi\gamma}{3}=1.47$~N/m for the  measured honeydew surface tension, $\gamma=44$~mN/m~\cite{liquidmarble2018pierre}. The droplet mass is estimated as $m_d\simeq(\rho_w\pi D^3/6)$ = 1.87$\times10^{-7}$~kg, using $D=0.71$~mm and  $\rho_w=1000$~kg/m$^3$. The corresponding droplet frequency is $f_o=1/2\pi\sqrt{64\gamma/ \rho_w D^3}\approx446\ \mathrm{s}^{-1}$. The  stylus frequency is $f=1/2\pi\sqrt{k_s/(m_s+m_d)}$, where  $m_s$ is the stylus mass and $k_s$ is the spring constant of its basal elastic element. The parameter values used in these calculations are summarized in Table~S4.  Viscous dissipation in the attached droplet is modeled by the damping ratio  $\zeta=c/2 \sqrt{k_d m_d}$, which is varied from $\zeta=0$  to $\zeta=0.75$ generate the theoretical response curves.

\subsection*{Capillary rectification model for nymphal SLF}
To quantify the nymphal mechanism, we consider a droplet confined between the two closing lobes of the split stylus and resolve the forces acting along the direction of release. The model is 2D and asks whether curvature asymmetry during stylus closure is sufficient to generate a positive net force toward detachment.

The driving term arises from the Young-Laplace pressure difference across the asymmetric droplet. If the the inner and outer interfaces have curvatures $\kappa_i$ and $\kappa_o$, the corresponding pressure difference is $ \Delta P = 2\gamma(\kappa_i-\kappa_o)$, and the resulting capillary driving force is
$ F_L = 2\gamma(\kappa_i-\kappa_o)A$, where $A$ is the projected droplet area, approximated as $A=\pi D^2/4$. Curvatures are measured directly from high-speed images.

The resisting forces are adhesion and viscosity. Adhesion from dynamic contact angle hysteresis is $ F_a=\gamma W(\cos\theta_r-\cos\theta_a)=\gamma W\,CAH_D,$
where $W$ is the effective channel width and $CAH_D$ is the dynamic contact angle hysteresis. Viscous resistance is estimated as $
F_v=\frac{\mu W l_d v}{\theta x},$
following the formulation of Reyssat and Qu\'er\'e~\cite{capillaryratchet2008manu}, where $l_d$ is the droplet length on the stylus, $v$ is droplet speed, $x$ is the distance from the apex, and $\theta$ is the stylus angle.

The net force along the stylus is therefore
\begin{align}
F_{\mathrm{net}}
=
2\gamma(\kappa_i-\kappa_o)A
-\gamma W\,CAH_D
-\frac{\mu W l_d v}{\theta x}.
\end{align}

A positive value of $F_{\mathrm{net}}$ indicates that the droplet is driven toward release. Rearranging the expression gives the condition
\begin{align}
\theta >
\frac{\mu W l_d V}
{x\gamma\left(2\Delta\kappa A-W\,CAH_D\right)},
\qquad
\Delta\kappa=\kappa_i-\kappa_o,
\end{align}
for $F_{\mathrm{net}}>0$.

The representative parameters to derive the theoretical $F_\textit{net}=0$ line are $\mu=1.55$~mPa$\cdot$s, $W\sim200~\mu$m, $l_d\sim0.2$~mm, $V\sim0.4$~m/s, $x\sim0.5$~mm, $\gamma=0.044$~N/m, $\Delta\kappa\sim1$~mm$^{-1}$, $D=0.61$~mm, $CAH_D\sim0.2$, and $\theta\sim50^\circ$. For the adult reference geometry used in Figure~5I, we use $\mu=1.55$~mPa$\cdot$s, $W\sim200~\mu$m, $l_d\sim1$~mm, $V\sim0.4$~m/s, $x\sim0.5$~mm, $\gamma=0.044$~N/m, $\Delta\kappa\sim0$~mm$^{-1}$, $D=0.71$~mm, $CAH_D\sim0.2$, and $\theta\sim100^\circ$.

\subsection*{Physical limits in rotational dynamics of organisms}
To estimate the fracture-limited rotational performance, we consider the stresses in a rotating appendage of characteristic length $L$. The internal tensile stress due to centrifugal force scales as $\sigma \sim \rho\,\omega^{2}L^{2}$, where $\rho$ is the material density and $\omega$ is the angular velocity. Imposing a material strength limit $\sigma \le \sigma_{\max}$ yields the velocity limit:
\begin{equation}
\omega_{\max} \sim \left(\frac{\sigma_{\max}}{\rho\,L^{2}}\right)^{1/2}.
\end{equation}

The angular acceleration is defined as $\alpha \equiv \tau/I$, where $\tau$ is the applied torque and $I$ is the mass moment of inertia about the rotation axis.  We use centrifugal tension to bound $\omega_\textit{max}$, while $\alpha_\textit{max}$ is bounded by the peak bending moment at the joint generated by actuator torque during rapid acceleration. For bending-dominated loading, the maximum bending moment at the joint before failure is $\tau_{\max} \sim \sigma_{\max} Z$, where $Z$ is the section modulus at the joint cross-section, which represents the appendage's resistance to bending. Thus, the acceleration limit is:
\begin{equation}
\alpha_{\max} = \frac{\tau_{\max}}{I} = \frac{\sigma_{\max} Z}{I}.
\end{equation}

Combining the stress-limited velocity and acceleration bounds yields a scale-invariant geometric constraint:
\begin{equation}
\alpha_{\max} = K\,\omega_{\max}^{2}, \qquad
K \sim \frac{\rho\,L^{2} Z}{I},
\end{equation}
where $K$ is a dimensionless geometric factor determined by the appendage's shape and mass distribution. Typically, $K$ spans from 10$^{-1}$ to 10$^{1}$ for highly optimized biological structures with an upper physical ceiling of O(10$^2$).

In addition to stress-limited bounds, rotational performance is constrained by the finite speed of elastic waves, which propagate at a characteristic speed $c_s\sim\sqrt{E/\rho}$, where $E$ is an elastic modulus \cite{physicallimit_2005_jan}. This implies a characteristic propagation time for a torque perturbation to be communicated along an appendage of length $L$, $t_{\mathrm{prop}}\sim L/c_s$. Requiring mechanically coherent rotation along the appendage ($\Delta t \geq t_\textit{prop}$), where $\Delta t$ denotes the characteristic timescale over which the torque (or acceleration) is applied, then yields an upper bound on angular speed,
$\omega_{wave}\sim c_s/L$ and $\omega_{max}\leq \omega_{wave}$. A corresponding acceleration bound follows by taking $\Delta\omega\sim\omega_{wave}$ (rest-to-peak) over the propagation time:
$\alpha_{wave}\sim \Delta\omega/t_{\mathrm{prop}}\sim (c_s/L)/(L/c_s)=c_s^2/L^2\sim E/(\rho L^2)$, thus implying $\alpha_{wave}\sim\omega_{wave}^2$. Mechanical coherence requires the torque application time to satisfy $\Delta t \gtrsim L/c_s$. 
Using $\Delta t\sim \Delta\omega/\alpha_{\max}$ with $\Delta\omega\sim\omega_{\max}$ and $\alpha_{\max}=K\omega_{\max}^2$ yields an elastic-wave constraint on the geometric factor,
\begin{equation}
K \lesssim \frac{c_s}{L\omega_{\max}}.
\end{equation}
Representative biological parameters place this ceiling at $K\sim O(10^2)$.

\subsection*{Fluctuations in honeydew droplet area and diameter of nymphal and adult SLF}
The normalized effective diameter, $D_\textit{eff,t}/D_\textit{eff,i}$, indicating how much the honeydew droplet deforms over time during droplet fall-off ranges from 0.816 to 1.140 with a mean value of 0.994$\pm$0.059 for adult SLF, where $D_\textit{eff,i}$ is the initial effective diameter (Figure S2). The extent of the fluctuations in the effective diameter, defined as $\sqrt{(D_\textit{eff,t}/D_\textit{eff,i}-1)^2}$, has a mean value of 0.043$\pm$0.040 (Figure S2). 
The values of $D_\textit{eff,t}/D_\textit{eff,i}$ of nymphal SLF are in the range between 0.96 and 1.11 with a mean value of 1.03$\pm$0.03 (Figure S2). The values of $\sqrt{(D_\textit{eff,t}/D_\textit{eff,i}-1)^2}$ have a mean value of 0.03$\pm$0.02 for nymphal SLF, representing 28$\%$ less fluctuation in the effective diameter compared to honeydew droplets from adult SLF (Figure S2). 


\subsection*{Detailed derivation of the geometric factor and physical limits}
For a baseline, non-optimized appendage, we approximate the rotating structure as a uniform solid cylinder of length $L$ and radius $R$ rotating about a proximal joint. The physical parameters are defined as follows: $m = \rho \pi R^2 L$ (mass), $I = \frac{1}{3} m L^2 = \frac{1}{3} \rho \pi R^2 L^3$ (mass moment of inertia), and $Z = \frac{\pi}{4} R^3$ (section modulus, bending). The dimensionless geometric factor $K$ is derived by combining the strength-limited acceleration $\alpha_{\max} \sim \sigma_{\max} Z / I$ and the velocity limit $\omega_{\max}^2 \sim \sigma_{\max} / (\rho L^2)$:
\begin{equation}K = \frac{\alpha_{\max}}{\omega_{\max}^2} = \frac{\sigma_{\max} Z / I}{\sigma_{\max} / (\rho L^2)} = \frac{\rho L^2 Z}{I}\end{equation}
Substituting the expressions for $Z$ and $I$ for a solid cylinder:
\begin{equation}K = \frac{\rho L^2 (\frac{\pi}{4} R^3)}{\frac{1}{3} \rho \pi R^2 L^3} = \frac{3}{4} \left( \frac{R}{L} \right)\end{equation}
For a typical slender biological appendage where the aspect ratio $R/L$ ranges from 0.01 to 0.1, we obtain $K \sim 10^{-2}$ to $10^{-1}$. This establishes the lower-bound baseline for biological rotation.

For morphologically optimized biological systems, the factor $K$ increases into the core biological band ($10^{-1} \lesssim K \lesssim 10^1$) through structural strategies, such as hollowing and mass concentration. Assuming that the structure fails by material yield rather than local buckling, for a hollow tube with fixed outer radius $R_o$ and inner radius $R_i = k R_o$ ($0 < k < 1$), the mass $m$ and section modulus $Z$ scale with $k$ as $m_{\mathrm{tube}}=\rho\pi R_o^2L(1-k^2)$ and $Z_{\mathrm{tube}}= I_a/R_o=\pi R_o^3(1-k^4)/4$, where the second moment of area about the neutral axis is $I_a=\pi(R_o^4-R_i^4)/4=\pi R_o^4(1-k^4)/4$. The ratio $Z/m$, and consequently $K$, increases by a factor of:
\begin{equation}
\frac{K_\textit{tube}}{K_\textit{solid}} = \frac{\frac{3R_o(1+k^2)}{4L}}{\frac{3R_o}{4L}} = (1+k^2)
\end{equation}
when $R=R_o$ for the solid cylinder. This allows for a higher $K$ by reducing the mass more than the structural resistance. As $k\to 1$, hollowing alone increases $K$ by at most a factor of 2. Therefore, much larger gains (toward $K\sim O(10)$) are more plausibly achieved by reducing inertia via mass concentration. 

Alternatively, concentrating mass by placing heavy muscles or joints near the axis of rotation reduces the dimensionless inertia $\tilde{I} \equiv I/mL^2=(r_g/L)^2$ when $I=mr_g^2$ where $r_g$ is the radius of gyration, leading to the expression for $K$ as: 
\begin{equation}
K= \frac{\rho L^2Z}{I}=\frac{\rho L^2 Z}{\tilde{I}mL^2}=\frac{\rho Z}{\tilde{I}m} 
\end{equation}
For a uniform rod rotating about its proximal end,
\begin{equation}
I=\int_{0}^{L} x^{2}\left(\frac{m}{L}\right)\,dx=\frac{1}{3}mL^{2},
\end{equation}
and thus
$\tilde{I}\equiv I/mL^{2}$=$1/3\approx$ 0.33. By approximating the concentrated mass as a point mass located at a distance $a$ from the joint, $\tilde{I}$ becomes $\tilde{I}=(r_g/L)^2\sim(a/L)^2$. Hence, if $a/L\approx0.2$, $\tilde{I}$ can drop to $\sim 0.05$. Since $\alpha_{\max}=\tau_{\max}/I$ (with $\tau_{\max}\sim \sigma_{\max}Z$), we have $\alpha_{\max}\sim \sigma_{\max}Z/I \propto 1/I \propto 1/\tilde{I}$ for fixed $m$ and geometry. Since $\alpha_{\max} \propto 1/\tilde{I}$, this optimization can increase $K$ by a factor of $\tilde{I}_{\text{rod}}/\tilde{I}_{\text{conc}}\approx 0.33/0.05\approx 6.6$ without requiring an increase in material strength or section modulus.

The absolute upper bound on rotational performance is dictated by the speed of elastic waves $c_s = \sqrt{E/\rho}$ \cite{physicallimit_2005_jan}. As established in the main text, mechanical coherence requires that the acceleration duration $\Delta t$ be no shorter than the wave propagation time, $t_{\mathrm{wave}} \sim L/c_s$.  Using $\alpha_{\max}=K\omega_{\max}^2$ and $\Delta t\sim \Delta\omega/\alpha_{\max}$ with $\Delta\omega\sim\omega_{\max}$ (rest-to-peak) gives $\Delta t\sim 1/(K\omega_{\max})$:
\begin{equation}\Delta t \sim \frac{1}{K \omega_{\max}} \gtrsim \frac{L}{c_s} \implies K \lesssim \frac{c_s}{L \omega_{\max}}
\end{equation}

To estimate this physical ceiling, we use representative values for high-performance biological systems such as chitin-based biological composites (i.e., sclerotized insect cuticle \cite{chitinproperty_2004_julian}) with $E\sim 1-20$ GPa, $\rho \sim 1000-1300$ kg/m$^3$, $c_s \approx 3000$ m/s ($E$=10 GPa, $\rho$=1200 kg/m$^3$), $L \approx 1$ mm ($10^{-3}$ m), and $\omega_{\max} \approx 10^4$ rad/s. Substituting these into the inequality yields:
\begin{equation}K \lesssim \frac{3000}{10^{-3} \times 10^4} = 300 \sim O(10^2)\end{equation}
This estimate suggests that $K\sim O(10^2)$ represents a hard physical ceiling for biological motion, as pushing beyond this regime would violate mechanical coherence and promote localized failure or instability before stresses can equilibrate along the appendage.

\clearpage




\setcounter{figure}{0}
\renewcommand{\thefigure}{S\arabic{figure}}

\begin{figure}[htbp]
\centering
\includegraphics[scale=0.5]{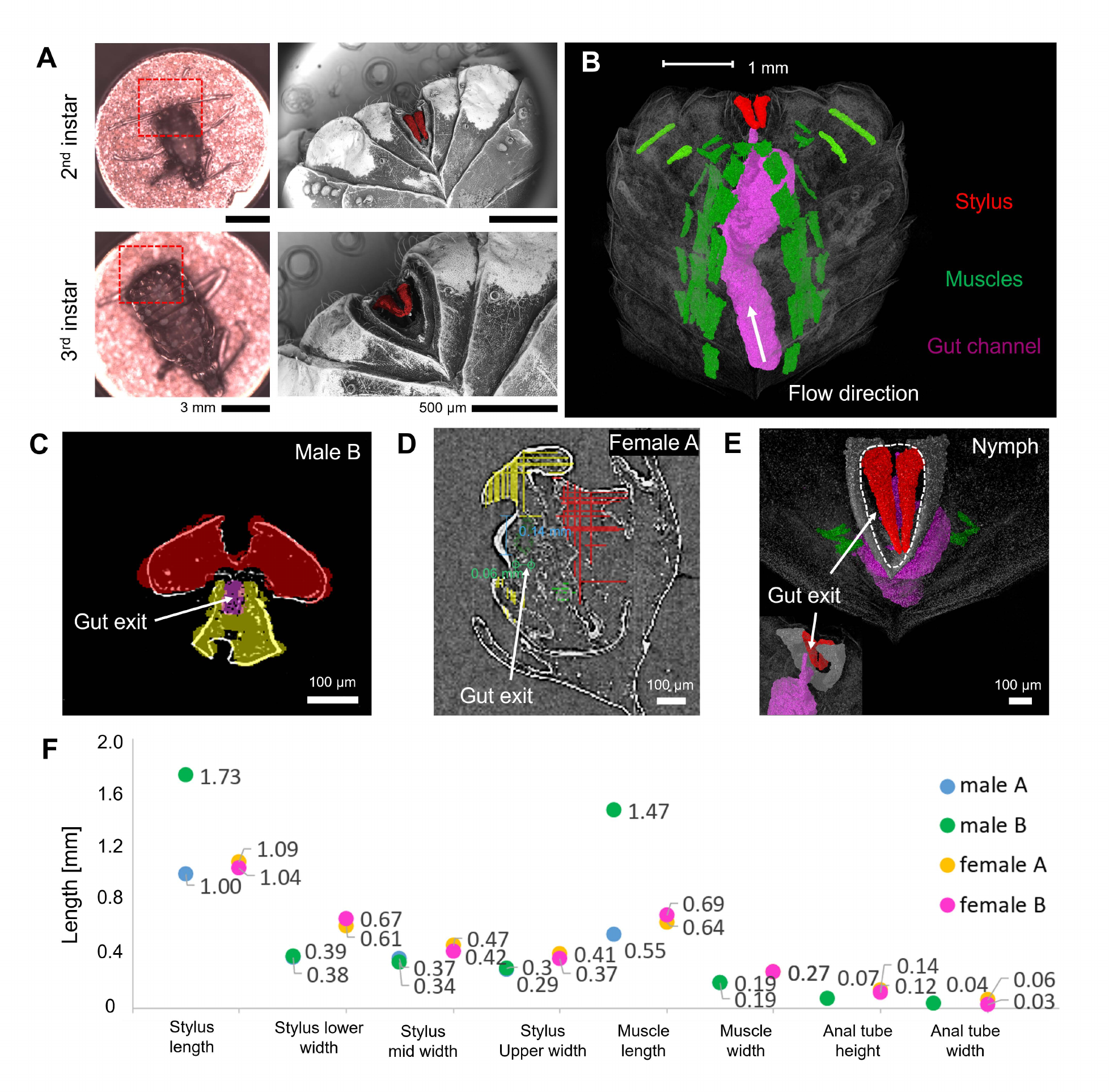}
\caption{Stylus and surrounding appendages of nymphal and adult SLF. (A) Microscopic and SEM images of 2nd instar and 3rd instar nymphal SLF. The ventral surface of the SLF is attached to the substrate. Red boxes indicate the location of the stylus. The stylus is colored red. (B) Micro-CT scan of 4th instar nymphal SLF. Muscle bundles near the dorsal surface and leg region are visualized. (C-E) Gut exits of (C,D) adult SLF (C: male; D: female SLF) and (E) 4th instar nymphal SLF. (F) Measurement of the length scales of the stylus and surrounding appendages of adult SLF. Two male and two female adult SLF were scanned.}
\end{figure}

\begin{figure}[htbp]
\centering
\includegraphics[scale=0.5]{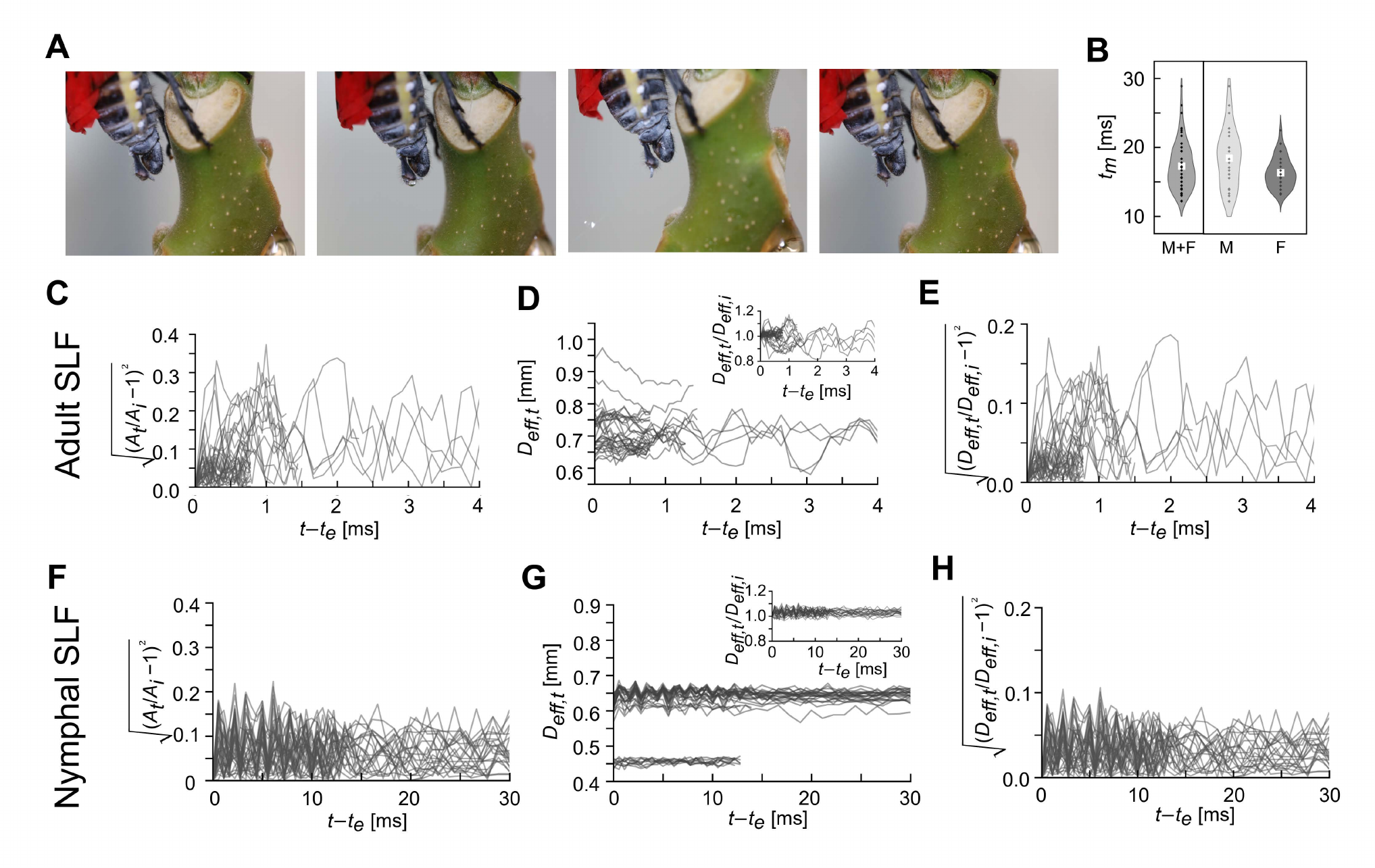}
\caption{Honeydew ejection of adult and nymphal SLF. (A) Sequential photos of liquid waste (honeydew) ejected from adult SLF. (B) Metastable time ($t_m$) for adult SLF (n=5, N=66) during the second droplet growth phase. Both male (n=2, N=29) and female SLF (n=3, N=37) show $t_m$ values around 15--20 ms. Error bars represent
the average value ± one standard deviation. White dots indicate the mean value. (C-H) Temporal variations in the surface area and effective diameter of honeydew ejected from (C-E) adult and (F-H) nymphal SLF. (C,F) Fluctuations in
the normalized surface area ($\sqrt{(A_t/A_i-1)^2}$). (D,G) Temporal variations in the effective diameter of ejected honeydew
droplets. (E,H) Fluctuations in the normalized effective diameter ($\sqrt{(D_\textit{eff,t}/D_\textit{eff,i}-1)^2}$). Insets show variations in the normalized surface area and effective diameter of the honeydew droplets, respectively.}
\end{figure}

\begin{figure}[htbp]
\centering
\includegraphics[scale=0.9]{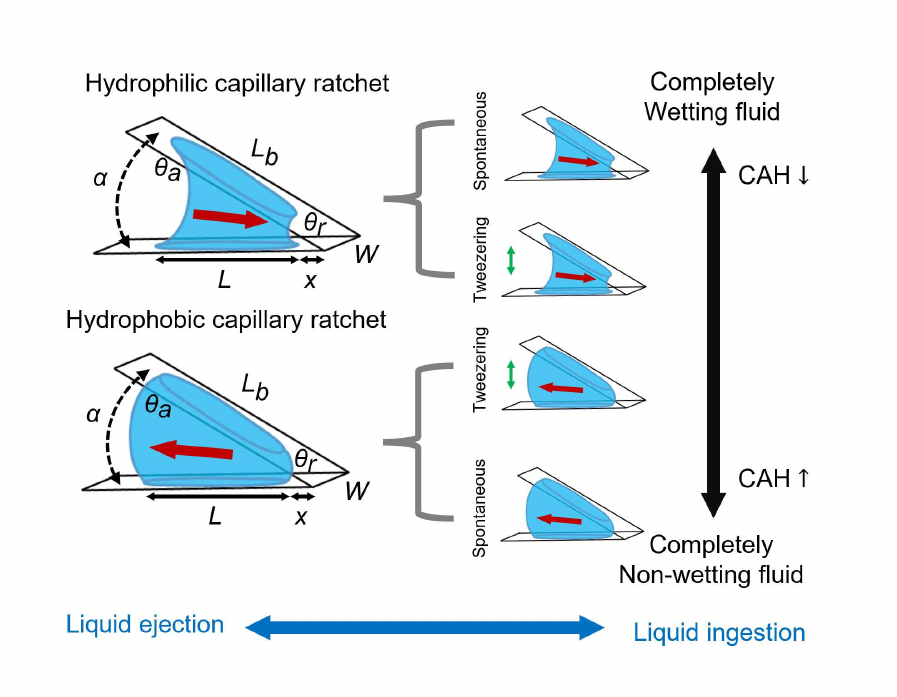}
\caption{Illustrations of hydrophilic and hydrophobic capillary ratchets for liquid ingestion or ejection. Four types of capillary ratchets are described according to wettability
from completely wetting fluids (very low CAH) to completely non-wetting fluids (very high CAH).
Completely wetting/non-wetting fluids can spontaneously move while partially wetting/non-wetting
fluids require a tweezering motion to drive the liquid transport.}
\end{figure}

\begin{figure}[htbp]
\centering
\includegraphics[scale=0.6]{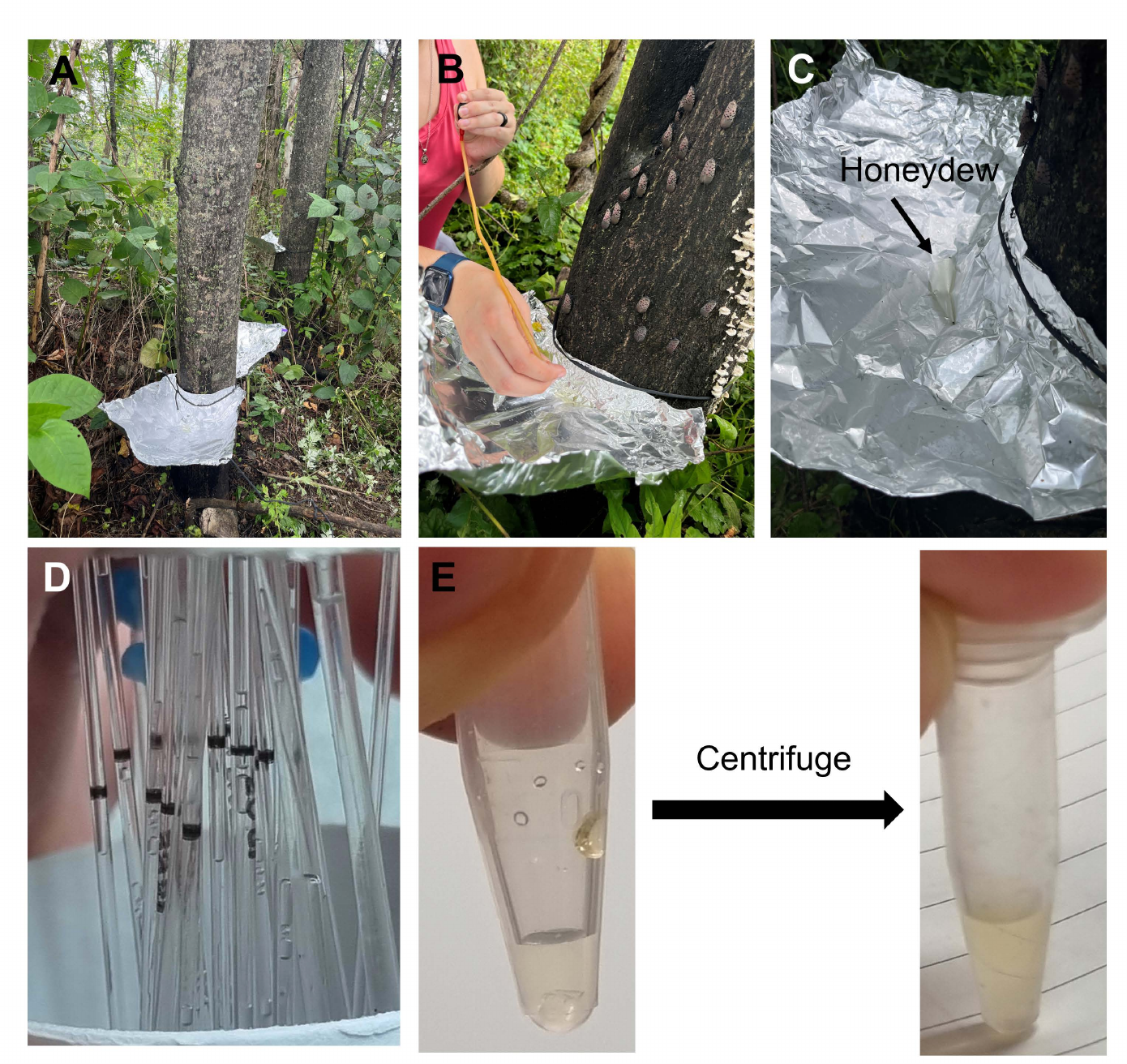}
\caption{Photographs of honeydew collection, storage, and the centrifugation process to measure its viscosity. (A-C) Collection of fresh honeydew beneath an Ailanthus tree in Stroudsburg, PA, using aluminum foil between 1:00 pm and 2:00 pm. (D) The collected honeydew was stored in wax-sealed capillary tubes to prevent evaporation and compositional changes. (E) As the amount of honeydew was too small, a short centrifugation step was performed for several seconds to detach the honeydew from the vial wall and maximize the amount of collected honeydew.}
\end{figure}

\begin{figure}[htbp]
\centering
\includegraphics[scale=0.5]{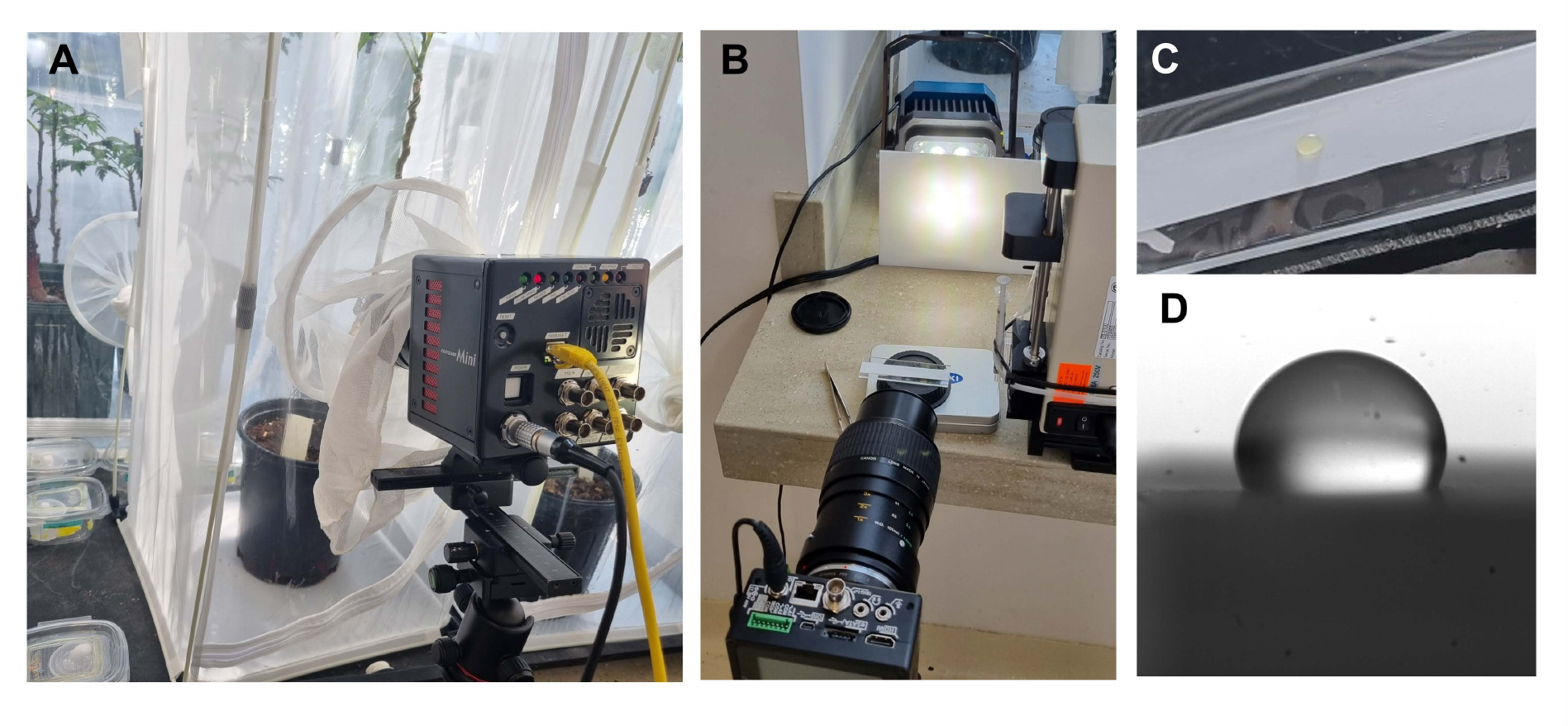}
\caption{Imaging setup. (A) Experimental setup for high-speed imaging of honeydew ejection in plant chambers. (B) A camera setup for capturing the contact angle (CA) and contact angle hysteresis of honeydew. (C) A honeydew droplet on the Teflon surface. (D) A side view of the droplet showing the CA value of the droplet on the Teflon surface.}
\end{figure}

\FloatBarrier
\clearpage
\setcounter{table}{0}
\renewcommand{\thetable}{S\arabic{table}}

\begin{table}[htbp]
  \centering
  \resizebox{\columnwidth}{!}{%
    \begin{tabular}{lcccc}
      \toprule
      Parameter & Female adult SLF & Male adult SLF & Adult SLF total & Nymphal SLF \\
      \midrule
      Max linear speed of stylus, $V_{s,\max}$ [m/s] 
      & (3,37) 
      & (3,33) 
      & (6,70) 
      & (3,36) \\

      Droplet ejection speed, $V_d$ [m/s] 
      & (3,28) 
      & (5,51) 
      & (8,79) 
      & (3,36) \\

      Speed ratio, $\lambda = V_d/V_s$ 
      & (3,28) 
      & (3,30) 
      & (6,58) 
      & (3,36) \\

      Droplet surface area, $A_t$ [mm$^2$] 
      & N/A 
      & N/A 
      & (6,34) 
      & (3,36) \\

      Initial effective diameter, $D_{\mathrm{eff},i}$ [mm] 
      & N/A 
      & N/A 
      & (6,34) 
      & (3,36) \\

      Droplet growth time, $\Delta t$ [ms] 
      & N/A 
      & N/A 
      & (6,34) 
      & (2,25) \\

    Post-launch droplet trajectory, $(X_m,Y_m)$ [mm,mm] 
      & N/A 
      & N/A 
      & (7,37) 
      & (3,36) \\

    Ejection velocity, $V_d$ [m/s] 
      & N/A 
      & N/A 
      & (8,79) 
      & (3,36) \\
      \bottomrule
    \end{tabular}
  }
  \caption{Summary of sample sizes ($n$, $N$) used for stylus kinematics and droplet measurements in female adult, male adult, pooled adult, and nymphal spotted lanternflies. Here, $n$ denotes the number of individual insects and $N$ denotes the number of droplet ejection events or droplets analyzed.}
  \label{tab:dataset_summary_slf}
\end{table}

\clearpage
\begin{table}[htbp]
  \centering
  \resizebox{\columnwidth}{!}{%
    \begin{tabular}{cccccccc}
      \toprule
      Sample & $\phi^{\text{a}}$ [$^\circ$] & $\theta_a^{\text{b}}$ [$^\circ$] & $\theta_r^{\text{c}}$ [$^\circ$] & $CAH^{\text{d}}$ & $\theta^{\text{e}}$ [rad] & $L_\text{CL}^{\text{f}}$ [mm] & $\gamma^{\text{g}}$ [mN/m] \\
      \midrule
      1 & 27.9 & 79.0 & 76.4 & 0.045 & 1.356 & 1.381 & 42.375 \\
      2 & 18.6 & 61.4 & 58.4 & 0.046 & 1.041 & 1.759 & 41.791 \\
      3 & 37.1 & 78.5 & 75.3 & 0.054 & 1.342 & 1.375 & 44.665 \\
      4 & 37.0 & 81.3 & 76.5 & 0.082 & 1.377 & 1.574 & 39.599 \\
      5 & 33.0 & 91.7 & 85.3 & 0.112 & 1.544 & 1.991 & 48.885 \\
      \bottomrule
    \end{tabular}
  }
  \caption*{\small
    \textsuperscript{a}Tilting angle. \ 
    \textsuperscript{b}Advancing contact angle. \ 
    \textsuperscript{c}Receding contact angle. \ 
    \textsuperscript{d}Static contact angle hysteresis. \ 
    \textsuperscript{e}Average of the advancing and receding contact angles. \
    \textsuperscript{f}Contact line length. \ 
    \textsuperscript{g}Surface tension of honeydew.
  }
  \caption{Measurement of contact angles and contact line length, and calculation of surface tension of honeydew ejected from spotted lanternflies.}
  \label{table1}
\end{table}

\clearpage
\begin{table}[htbp]
  \centering
  \resizebox{0.7\columnwidth}{!}{%
    \begin{tabular}{cc}
      \toprule
      Sample & Dynamic viscosity, $\mu$ [mPa$\cdot$s] \\
      \midrule
      1 & 1.476 \\
      2 & 1.323 \\
      3 & 1.331 \\
      4 & 1.994 \\
      5 & 1.880 \\
      6 & 1.279 \\
      \bottomrule
    \end{tabular}
  }
  \caption{Measurement of dynamic viscosity of honeydew ejected from spotted lanternflies.}
  \label{table2}
\end{table}

\clearpage
\begin{table}[htbp]
  \centering
  \resizebox{\textwidth}{!}{%
    \begin{tabular}{@{} c c c c c @{}}
      \toprule
      \textbf{Parameter} & \textbf{Description} & \textbf{Adult SLF} & \textbf{Nymphal SLF} & \textbf{Unit}  \\
      \cmidrule(lr){1-5}
      $m_d$ & Droplet mass & $1.87\times 10^{-7}$ & $1.19\times 10^{-7}$ & kg \\
      $k_d$ & Spring constant of droplet & \multicolumn{2}{c}{1.47} & N/m \\
      $f_o$ & Undamped natural frequency of droplet & 446 & 561 & 1/s \\
      $f$ & Frequency of lower spring & $579\pm382$ & $856\pm391$ & 1/s \\
      $f_o/f$ & Frequency ratio & 0.79$\pm$0.39 & 0.77$\pm$0.23 & — \\
      $\lambda$ & Speed ratio of droplet and stylus & 0.62$\pm$0.11 & 0.18$\pm$0.04 & — \\
      \bottomrule
    \end{tabular}
  }
  \caption{Summary of the two-spring model and parameters for the respective adult and nymphal spotted lanternflies.}
  \label{table6}
\end{table}

\clearpage
\begin{table}[htbp]
  \centering
  \resizebox{\textwidth}{!}{%
    \begin{tabular}{cccccccccccc}
      \toprule
      System & Rotational appendage & Classification & Surrounding fluids & ${\Omega_\text{max}}^{\text{a}}$ [$^\circ$/s] & $\dot{\Omega_\text{max}}^{\text{b}}$ [$^\circ$/s$^2$] & ${\Omega_\text{max}}^{\text{a}}$ [rad/s] & SD$^{\text{ c}}$ [rad/s] & $\dot{\Omega_\text{max}}^{\text{b}}$ [rad/s$^2$] & SD$^{\text{ c}}$ [rad/s$^2$] & Refs \\
      \midrule
      Trap jaw ant ($\textit{O. brunneus}$) & Jaw & LaMSA & Air & 2.53E+06 & 4.08E+10 & 4.42E+04 & 4.50E+03 & 7.12E+08 & 1.17E+08 & \cite{trap_jaw_ant_2022_gregory} \\
      Trap jaw ant ($\textit{M. camillae}$, major work caste) & Jaw & LaMSA & Air & 3.21E+06 & 3.44E+11 & 5.60E+04 & 8.70E+03 & 6.00E+09 & 2.10E+09 & \cite{trap_jaw_ant_2018_fredrick} \\
      Trap jaw ant ($\textit{M. camillae}$, minor work caste) & Jaw & LaMSA & Air & 6.36E+06 & 9.85E+11 & 1.11E+05 & 1.84E+04 &  1.72E+10 & 2.70E+09 & \cite{trap_jaw_ant_2018_fredrick} \\
       Springtail & Furca & LaMSA & Air & 2.11E+05 & 2.04E+09 & 3.69E+03 & 4.59E+02 & 3.55E+07 & 3.27E+06 & \cite{springtail2022victor, springtail2024jacob} \\
    Female adult SLF & Stylus & Elastic catapult & Air, Honeydew & 9.43E+04 & 5.93E+08 & 1.65E+03 & 1.68E+02 & 1.04E+07 & 2.99E+06 &This study \\
      Male adult SLF & Stylus & Elastic catapult & Air, Honeydew & 6.99E+04 & 3.53E+08 & 1.22E+03 & 2.17E+02 & 6.17E+06 & 2.93E+06 &This study  \\
      Nymphal SLF & Stylus & Capillary ratchet & Air, Honeydew & 1.01E+05 & 1.19E+09 & 1.76E+03 & 3.63E+02 & 1.68E+07 & 5.33E+06 & This study  \\
      Mantis shrimp ($\textit{N. bredini}$) & Link b-c & LaMSA & Water & 9.08E+04 & 3.32E+08 & 1.59E+03 & 7.55E+02 & 5.80E+06 & N/A & \cite{mantis_shrimp_2021_emma} \\
      Mantis shrimp ($\textit{O. scyllarus}$) & Link b-c & LaMSA & Water & 3.47E+04 & 5.39E+07 & 6.05E+02 & 5.10E+01 & 9.40E+05 & N/A & \cite{mantis_shrimp_2021_emma} \\
      Mantis shrimp ($\textit{N. bredini}$) & Link b-c & LaMSA & Water & 1.62E+05 & 6.30E+08 & 2.82E+03 & 6.82E+02 & 1.10E+07 & N/A & \cite{mantis_shrimp_2021_emma} \\
    Mantis shrimp-inspired robot & Link b-c & LaMSA & Water & 1.76E+04 & 2.52E+06 & 3.07E+02 & 1.48E+02 & 4.40E+04 & N/A & \cite{mantis_shrimp_2021_emma} \\
    Mantis shrimp-inspired robot & Link b-c & LaMSA & Air & 1.07E+05 & 7.45E+07 & 1.86E+03 & 7.80E+01 & 1.30E+06 & N/A & \cite{mantis_shrimp_2021_emma} \\
      Fruit fly & Wing & Muscle-driven stroke & Air & 8.02E+04 & 1.15E+08 & 1.40E+03 & N/A & 2.00E+06 & N/A & \cite{fruitfly2014florian, mosquitofruitfly2022wouter}  \\
      Malaria mosquito & Wing & Muscle-driven stroke & Air & 6.88E+04 & 2.58E+08 & 1.20E+03 & N/A & 4.50E+06 & N/A & \cite{mosquito2017florian, mosquitofruitfly2022wouter} \\
    Sharpshooter & Stylus & Elastic catapult & Air, Water & 3.31E+04 & 2.50E+07 & 5.78E+02 & 2.29E+02 & 4.36E+05 & N/A & \cite{sharpshooter2023elio,Fluidejection2024elio}  \\
    Hummingbird & Wing & Muscle-driven stroke & Air & 2.00E+04 & 5.50E+06 & 3.49E+02 & N/A & 9.60E+04 & N/A & \cite{hummingbird2018rivers}  \\
    Finger snap & Finger & Muscle-driven stroke & Air & 7.79E+03 & 1.60E+06 & 1.36E+02 & 2.40E+01 & 2.80E+04 & 5.00E+03 & \cite{finger_snap_2021_raghav}  \\
    Baseball player & Arm & Muscle-driven stroke & Air & 7.51E+03 & 5.00E+05 & 1.31E+02 & 1.99E+01 & 8.73E+03 & N/A & \cite{ballpitch1994glenn, ballpitch2006eric}  \\
      \bottomrule
    \end{tabular}
  }
  
  \caption*{\small
    \textsuperscript{a}Maximum angular velocity. \ 
    \textsuperscript{b}Maximum angular acceleration. \
    \textsuperscript{c}Standard deviation.
  }
    \caption{Comparison of maximum angular velocity and acceleration of various systems with ultrafast rotational motions.}
  \label{table4}
\end{table}

\FloatBarrier
\clearpage

\movie{Micro-CT scans of nymphal and adult SLF showing the differences in structural appendages related to honeydew excretion.}

\movie{High-speed video on honeydew droplet ejection using catapult-like stylus of adult SLF.}

\movie{High-speed video on post-launch honeydew droplet ejected by adult SLF.}

\movie{Sequential honeydew excretion by nymphal and adult SLF.}

\movie{Honeydew droplet ejection using hinge-like stylus of nymphal SLF.}

\dataset{Dataset\_S01.xlsx}{Raw data on the kinematics of the stylus in adult and nymphal SLF. Each tab includes Tab 1-3: female adult SLF (n=2,N=3), Tab 4-6: male adult SLF (n=2,N=3), and Tab 7-9: nymphal SLF (n=2,N=16).}

\dataset{Dataset\_S02.xlsx}{Raw data on the kinematics of droplets from adult and nymphal SLF. Each tab includes Tab 1-3: female adult SLF (n=3,N=6), Tab 4-7: male adult SLF (n=4,N=29), and Tab 8-10: nymphal SLF (n=3,N=28).}

\dataset{Dataset\_S03.xlsx}{Time scales during the honeydew excretion. Each tab includes Tab 1: droplet growth time scales of adult (n=6,N=64) and nymphal SLF (n=2,N=25) and the derived flow speed ($u$) through a gut exit and the resulting excretion energy ($\eta_e$), Tab 2: meta-stable time scales ($t_m$) of adult SLF, Tab 3: droplet ejection frequency of nymphal SLF, and Tab 4: the excretion and rest time scales of nymphal SLF.}

\dataset{Dataset\_S04.xlsx}{Trajectories of honeydew droplets ejected from adult and nymphal SLF.}

\dataset{Dataset\_S05.xlsx}{Fluctuations in surface area of ejected honeydew from adult and nymphal SLF.}

\dataset{Dataset\_S06.xlsx}{Stylus and honeydew ejection speed ratio of adult and nymphal SLF. Each tab includes Tab 1: maximum linear speed of stylus (female adult SLF: n=3, N=37, male adult SLF: n=3, N=33, nymphal SLF: n=3, N=36), Tab 2: honeydew ejection speed (female adult SLF: n=3, N=28, male adult SLF: n=5, N=51, nymphal SLF: n=3, N=36), and Tab 3: speed ratio (female adult SLF: n=3, N=28, male adult SLF: n=5, N=30, nymphal SLF: n=3, N=36).}

\dataset{Dataset\_S07.xlsx}{Dimensionless analysis parameters. Each tab includes Tab 1: the gut exit diameter ($d$) of adult and nymphal SLF, and Tab 2: the gut exit diameter ($d$), flow speed through the orifice ($u$), pre-launch dimensionless numbers including Bond number ($Bo_d$) and Weber number ($We_d$), droplet diameter ($D$), and post-launch dimensionless numbers including Bond number ($Bo_D$), Weber number ($We_D$), and spinning Weber number ($We_\Omega$) of SLF and other organisms. The data for other organisms were regenerated from \cite{Fluidejection2024elio}.}

\dataset{Dataset\_S08.xlsx}{Two-spring model parameters, including Tab 1: the initial effective diameter ($D_\textit{eff,i}$), the upper/lower spring frequency ratio ($f_o/f$), and speed ratio ($\lambda$) values of adult and nymphal SLF. Tab 2 includes the stylus frequency ($f$) values of adult SLF (n=4,N=13).}

\dataset{Dataset\_S09.xlsx}{Capillary ratchet model parameters, including Tab 1: inner ($\kappa_i$) and outer ($\kappa_o$) droplet curvatures, Tab 2: temporal variations in average and standard deviations of the inner and outer curvatures and the derived Laplace pressure force ($F_L$), Tab 3: dynamic contact angle hysteresis (CAH) and the derived adhesion force ($F_a$), Tab 4: temporal variations in average and standard deviations of the dynamic CAH and the $F_a$ values, Tab 5: $l_d$, $x$, and viscous forces ($F_v$), and Tab 6: temporal variations in average and standard deviations of the $l_d$, $x$, and $F_v$ values for nymphal SLF (n=2, N=26).}

\dataset{Dataset\_S10.xlsx}{Superfast rotational performance of the SLF stylus, including Tab 1: maximum angular velocity and acceleration data for the rotating stylus of adult (female: n=3, N=37, male: n=3, N=33) and nymphal (n=3, N=36) SLF, and Tab 2: maximum linear acceleration of the stylus in adult (female: n=3, N=37, male: n=3, N=33) and nymphal SLF (n=3, N=36).}

\printbibliography[title={Supporting References}]
\end{refsection} 

\end{document}